\documentclass{interact}

\usepackage{epstopdf}
\usepackage[caption=false]{subfig}

\usepackage{array}
\usepackage{wrapfig}

\newcolumntype{C}[1]{>{\centering\arraybackslash}p{#1}}

\usepackage[numbers,sort&compress]{natbib}
\bibpunct[, ]{[}{]}{,}{n}{,}{,}
\renewcommand\bibfont{\fontsize{10}{12}\selectfont}
\makeatletter
\def\NAT@def@citea{\def\@citea{\NAT@separator}}
\makeatother

\usepackage[hypertexnames=false]{hyperref}
\usepackage[dvipsnames]{xcolor} 
\hypersetup{
    colorlinks=true,
    linkcolor=NavyBlue,    
    citecolor=NavyBlue,    
    urlcolor=NavyBlue,
    filecolor=NavyBlue
}

\theoremstyle{plain}

\theoremstyle{definition}

\theoremstyle{remark}

\begin{document}

\articletype{Introductory review article}

\title{JWST provides a new view of cosmic dawn: latest developments in studies of early galaxies} 

\author{
\name{Jorryt Matthee\textsuperscript{*}\thanks{Email: jorryt@ista.ac.at} }
\affil{\textsuperscript{*}Institute of Science and Technology Austria (ISTA), Am Campus 1, 3400 Klosterneuburg, Austria}
}

\maketitle

\begin{abstract}
Studies of the distant Universe are providing key insights into our understanding of the formation of galaxies. The advent of the James Webb Space Telescope (JWST) has significantly enhanced our observational capabilities, leading to an expanded redshift frontier, providing unprecedented detail in the characterization of early galaxies and enabling the discovery of new populations of accreting black holes. This review aims to provide an introduction to the basic processes and components that shape the observed spectra of galaxies, with a focus on their relevance to techniques with which high-redshift galaxies are selected. The review further introduces specific topics that have attracted significant attention in recent literature, including the discovery of highly efficient galaxy formation in the early Universe, the relation between galaxies and the process of reionization, new insights into the formation of the first stars and the enrichment of interstellar gas with heavy elements, and breakthroughs in our understanding of the origins of supermassive black holes.

\end{abstract}


\section{Introduction}
\begin{figure*}[h]
    \centering
    \includegraphics[width=0.93\linewidth]{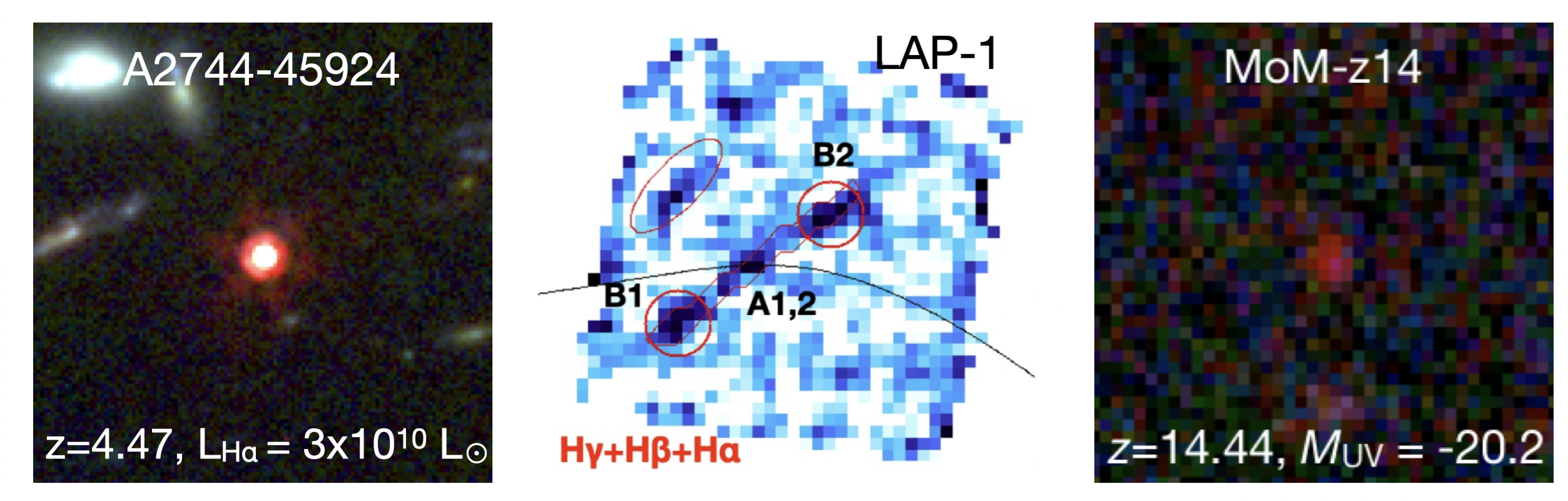}
    \caption{Images that showcase JWST's abilities to push the frontiers of extra-galactic astrophysics. On the left is a false-color image constructed from JWST/NIRCam images in the F070W/F200W/F356W filters, $5\times5$'' on the side. The central object is one of the most luminous examples of the newly discovered population of `little red dots' that are accreting supermassive black holes whose red colors are attributed to coverage by dense gas. In the middle is a $3\times3$'' JWST/NIRSpec pseudo-narrow band image of the Balmer emission lines of a highly magnified galaxy that is constructed from 3D spectroscopic data (image adapted from Vanzella et al. \citep{Vanzella23}). The galaxy is at a redshift of $z=6.6$ and has a mass comparable to a globular cluster and a metallicity less than 0.4 \% solar -- one of the most metal poor galaxies known. On the right is a zoomed in $1\times1''$ JWST/NIRCam image in the F090W/F115W/F277W filters of the most distant galaxy currently known at $z=14.4$ (image adapted from Naidu et al. \citep{Naidu25}), whose light has traveled 13.3 billion years to reach us. }  
    \label{fig:intro}
\end{figure*}

Observing and understanding the most distant light has been challenging human imagination and technical capabilities for centuries. Now, the unique infrared capabilities of the James Webb Space Telescope (JWST) are revolutionizing the study of the most distant galaxies \citep[e.g.][]{Adamo24}. Data from JWST have so-far allowed us to push the horizon at which we have observed stellar light from galaxies at redshifts $z\approx14$ \citep{Carniani24,Naidu25},  corresponding to a time of about 300 million year after the Big Bang, or a lookback time of about 13.3 billion years. A galaxy's redshift primarily originates from the expansion of the Universe that stretches the light of a photon since its emission. The redshift corresponds to the relative scales of the Universe between emission and now as $(1+z_{\rm em}) = \frac{a_{\rm now}}{a_{\rm em}}$. By combining the observed redshift of the light from a galaxy with cosmological models one can derive the lookback time. JWST data have also led to the identification of new abundant populations of active galactic nuclei \citep{Harikane23,Matthee24,Maiolino23b}, they enabled unprecedented characterization of the properties of interstellar gas, such as the chemical composition \citep{Vanzella23,Chemerynska24,Ji24}, and yielded new insights into how the light from early galaxies impacted the intergalactic medium in their large-scale environments \citep{Tang24,Witstok24,Kakiichi25,Kashino25}. Figure $\ref{fig:intro}$ shows zoomed-in images of JWST images of such a newly identified active galactic nucleus, one of the most primordial galaxies and the most distant galaxy known, respectively.

The field of \textit{extra-galactic} astrophysics emerged with the realization that many nebulae observed in the night sky are located far outside our own Milky Way galaxy \citep[e.g.][]{Slipher17,Hubble29}. The redshift of the majority of nebulae (galaxies) was among the first solid evidence for our current cosmological understanding of an expanding Universe. As galaxies signpost the peaks of matter over-densities, mapping the distribution of galaxies through space and time has been one of the cornerstones of cosmology. The formation of galaxies is determined by cosmological parameters, such as the matter density and the amplitude of temperature perturbations during the Big Bang. The detection of the Big Bang afterglow in the 1960s, the detailed characterization of its perturbations in the following decades and the maps of distributions and distances to galaxies have anchored a cosmological model, providing the boundary conditions for theories of galaxy formation (albeit with important poorly understood dark matter and dark energy).

Although the boundaries of the visible Universe have been reached with the detection of the cosmic microwave background, the detection of the first light from stars that formed in the Universe remains one of the key quests to date. The progress in the study of distant galaxies historically has trailed technological advances and can be exemplified by considering the evolution of the history of the known redshift record. Optical spectroscopy with photographic plates led to a gradual increase in the redshift record out to $z=0.2$ in the first half of the 20th century. The new radio domain that opened in the 1950's led to the discovery of quasars and a suite of new redshift records, as the spectral steepness at radio frequencies is a good redshift indicator, breaking the $z=2$ limit in 1965 \citep{Schmidt65} and out to $z\approx5$ by the 1990s. Quasars, which are luminous compact sources of light powered by the accretion of matter on a supermassive black hole, would remain the most distant known sources until the serendipitous detection of a pair of gravitationally lensed galaxies at $z=4.92$ in 1997 \citep{Franx97}. The development of the charge couple device and the generation of 10m-class optical telescopes, combined with the Hubble Space Telescope (HST), pushed the redshift record into the epoch of reionization, with redshifts confirmed at $z=6.5$ in 2002 \citep{Hu02} and $z=11.1$ in 2016 \citep{Oesch16}, although revised to $z=10.6$ by JWST \citep{Bunker23}, whereas some photometric candidates existed beyond $z>10$ \citep[e.g.][]{Finkelstein22}. Since 2022, JWST enables sensitive infrared imaging and spectroscopy from space, which has rapidly expanded the spectroscopic redshift record to $z=14.4$ \cite{Naidu25}. This record will very likely be broken, as there is no major technical limitation preventing us from spectroscopically confirming the redshift of a galaxy out to redshifts of $z\sim20$.%

\begin{figure*}[h]
\centering
    \includegraphics[width=12cm]{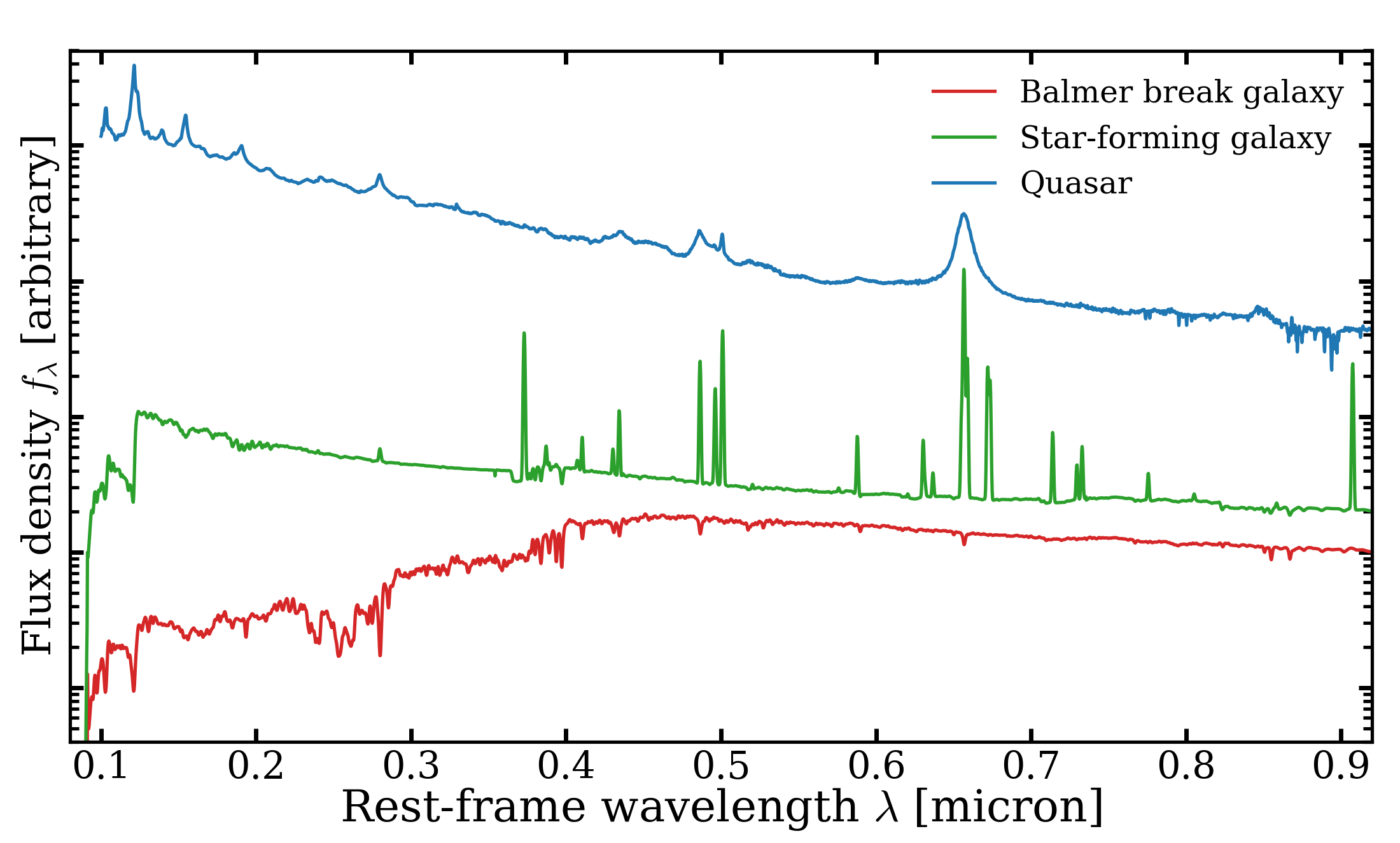}
    \caption{Example spectra of galaxies and quasars in the distant Universe. The two galaxy spectra in red and green are based on models \cite{Williams18}, while the quasar spectrum (blue) is an empirical template \cite{Selsing16}. The star-forming galaxy (green) has young stars and little dust attenuation and therefore a blue continuum with strong narrow emission-lines as [O{\sc ii}], H$\beta$, [O{\sc iii}] and H$\alpha$. The Balmer break galaxy (red) has an older stellar population and is not actively forming stars. Absorption line features from stellar atmospheres can be identified, as well as a break at the Balmer wavelength of $3.64$ $\mu$m. Note that attenuation due to neutral gas in the intergalactic medium is only applied at $\lambda<0.912 \mu$m. The quasar shows a power-law like blue continuum with various strong broad emission-lines as Lyman-$\alpha$, C{\sc iv}, Mg{\sc ii}, H$\beta$ and H$\alpha$. }\label{fig:sed}
\end{figure*}

\section{The spectrum of galaxies} \label{sec:spectrum} 

The physical information that we retrieve from galaxies predominantly relies on our interpretation of their spectrum\footnote{Important exceptions to this are measurements of the clustering of galaxies, the gravitational deflection of light from background sources, and measurements of absorption lines in the light from background sources. However, all those techniques require us to detect the light from background galaxies in the first place.}. 
The spectrum of galaxies is a composite of the light-emitting and obscuring components that include populations of stars, clouds of ionized gas around hot stars, accreting (supermassive) black holes and their surroundings and neutral gas and dust in the interstellar medium. Various detailed review articles have been written on models of the spectrum of galaxies, for example \cite{Conroy13,Iyer25}. How can we use light to tell the difference between a young and an old galaxy? What are the key spectral features that we use to find galaxies and measure their distance? How do we know that there is an active supermassive black hole in a galaxy? What does interstellar dust do to the observed light? How can we measure detailed properties of the intergalactic gas whose emitted light itself we cannot detect? 

Here, the main focus is on the rest-frame ultra-violet (UV) and optical spectrum of young galaxies (ages $<10^9$ yr). This is because our current telescopes can most sensitively probe galaxies in this wavelength range due to their UV brightness, but also because strong nebular emission lines are observed in this wavelength range. These include the hydrogen recombination lines from the Lyman and Balmer series (e.g. Lyman-$\alpha$ at $\lambda_0=1215.67$ {\AA}, and H$\alpha$ and H$\beta$ at $6564.6, 4862.7$ {\AA}, respectively) or collisionally excited lines from atoms as oxygen, e.g. [O{\sc ii}]$_{3727,3729}$ or [O{\sc iii}]$_{4960,5008}$. In the early Universe ($z>2$, the first 3 Gyrs), most of these emission-lines are redshifted into the near-infrared and are accessible with spectroscopic instruments onboard JWST. JWST has already measured such lines in galaxies with stellar masses as low as $10^6$ M$_{\odot}$ at redshifts $z\sim8$ \citep{Atek24} and out to redshift $z\sim12$ \citep{Zavala24}. 

Figure $\ref{fig:sed}$ shows three example spectra of different types of galaxies that are observed in the early Universe. These illustrate the typically flat and blue UV continuum and narrow emission-lines in the rest-frame optical spectrum of young star-forming galaxies, the relative smooth and featureless continuum of an inactive `quenched' galaxy and the broad lines in quasars. The star-forming galaxy also clearly displays the Lyman-$\alpha$ break (at 1.216 $\mu$m) and the Lyman-continuum break (at 0.912 $\mu$m), which are often used for the identification of distant galaxies (see \S $\ref{sec:find}$).

\subsection{Stellar emission} \label{sec:stars}
Stars are a key component of galaxies and are among the main contributors of the light that we observe from distant galaxies. Spectra of individual stars resemble black-body spectra that peak at the effective temperatures of their atmospheres. The temperature, which in turn sets the color, and the luminosity are primarily dependent on the mass and chemical composition of the star. It is commonly assumed that masses of stars in a population follow a universal initial mass distribution (the so-called initial mass function; IMF, see \cite{Hopkins18} for a review). The IMF is usually parameterized with a broken power law \citep{Kroupa01} or a log-normal \citep{Chabrier03} distribution with limits $\approx0.1-100$ M$_{\odot}$. The majority of the aggregate stellar mass is in relatively cool, long-lived low-mass stars, whereas the majority of the light is emitted by hot, short-lived massive stars. Stars with lower metallicities have somewhat hotter temperatures because of a reduced line-blanketing effect.

\begin{figure*}[h]
\centering
    \includegraphics[width=12cm]{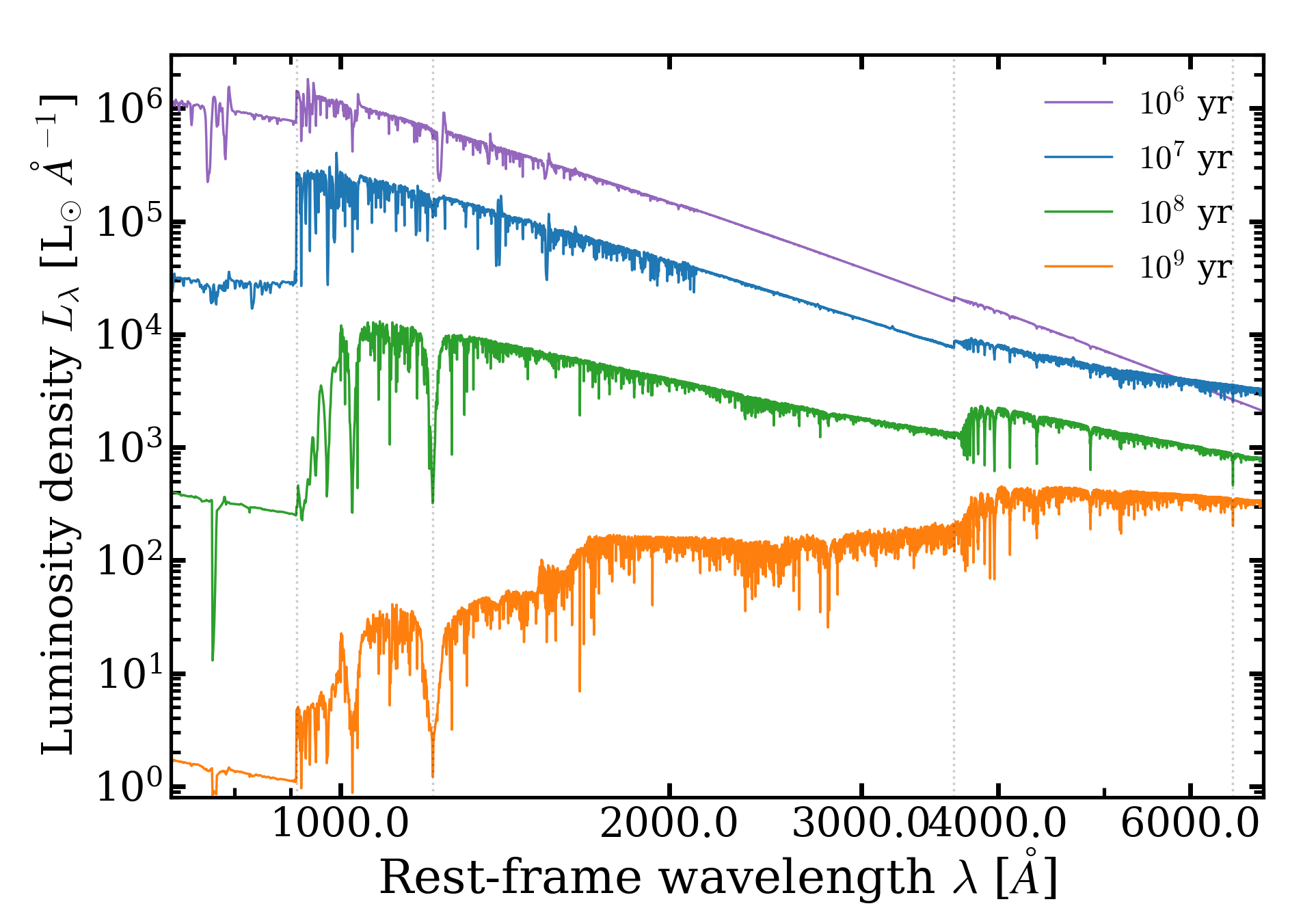}
    \caption{Model spectra of simple stellar populations with different ages in the rest-frame UV to optical. The models are constructed using the Binary Population and Spectral Synthesis code, version 2.3, and describe stellar populations with a mass $10^6$ M$_{\odot}$, a 20 \% solar metallicity and a broken power-law IMF (with slopes of $\alpha=-1.3, -2.35$ at masses below and above 0.5 M$_{\odot}$, respectively). Dotted vertical lines highlight key wavelengths of interest: the Lyman edge at 912 {\AA}, the Ly$\alpha$ transition at 1216 {\AA}, the Balmer edge at 3645 {\AA} and the H$\alpha$ transition at 6564 {\AA}. The models clearly illustrate the strong dependence of the mass-to-light ratio on age, the slight reddening of the SEDs with increasing age, the Lyman break and the development of the Balmer break around $10^8$ yr.}\label{fig:starsed}
\end{figure*}

The stellar emission from a galaxy is the composite of a population of stars with a range in masses, ages, and metallicities. The building blocks of most galaxy spectral energy distribution (SED) models are so-called single stellar population, which are typically populations of stars with a combined mass of $10^6$ M$_{\odot}$ that sample the IMF. Figure $\ref{fig:starsed}$ shows a spectral model that includes binary interactions. The metallicity is 20 \% the solar value which is typical for early galaxies as there was less time for metal enrichment. The stellar atmospheres have enhanced ratios of $\alpha$ elements as oxygen compared to the iron abundance based on measurements low metallicity stars and high-redshift galaxies \cite[e.g.][]{Steidel16}. This $\alpha$-enhancement is because early abundances are more strongly impacted by core-collapse supernova (which produce $\alpha$ elements) rather than Type Ia supernova (which produces more iron-rich gas). 

Figure $\ref{fig:starsed}$ shows various important spectral characteristics of stellar populations. First, it shows that the mass-to-light ratio varies strongly with age, with a UV luminosity that varies with a factor of a few in the first 10 Myr and about two orders of magnitude in the first 100 Myr. Second, the spectrum varies with age, and the fading of the SED with age is strongest in the UV. These two trends are driven by the short lifespans of the hottest massive stars. As more temperate A stars start to dominate the spectrum at ages $\sim100-500$ Myr, one can note strong Lyman-$\alpha$ absorption and strong Balmer breaks, as well as numerous absorption lines from the Balmer series lines. These absorption lines are associated with transitions in hydrogen atoms that are strongest in atmospheres with temperatures of $\sim10,000$ K. Other, somewhat less prominent, but important features are the numerous narrow absorption lines that are visible in the rest-frame UV at almost all ages. These features arise from metal lines in the photosphere (in particular iron). 

The key property that impacts the composite spectrum of a galaxy is the star formation history. These were usually parameterized as a constant star formation history or the relatively flexible delayed-$\tau$ models where the star formation rate rises and declines following a star formation rate SFR $\propto e^{-t/\tau}$. Recent developments include the so-called ‘nonparametric’ histories that split the stellar populations into various age bins with flexible continuity criteria (see \cite{Iyer25} for a detailed overview). However, these models become computationally expensive. Estimates of stellar mass are strongly sensitive to the choice of the star formation history. This is primarily because a massive old population can relatively easily be outshone by a recent burst, especially in the UV. While such methods are currently our best way of quantifying the uncertainties on the stellar mass (and other parameters), it is important to note that they may become template-limited, e.g. due to uncertainties in the contribution from accreting supermassive black holes, accurate models for massive stars at low metallicity, or assumptions in the geometry that impact the nebular emission and dust attenuation.

\begin{figure*}
\centering
    \includegraphics[width=14cm]{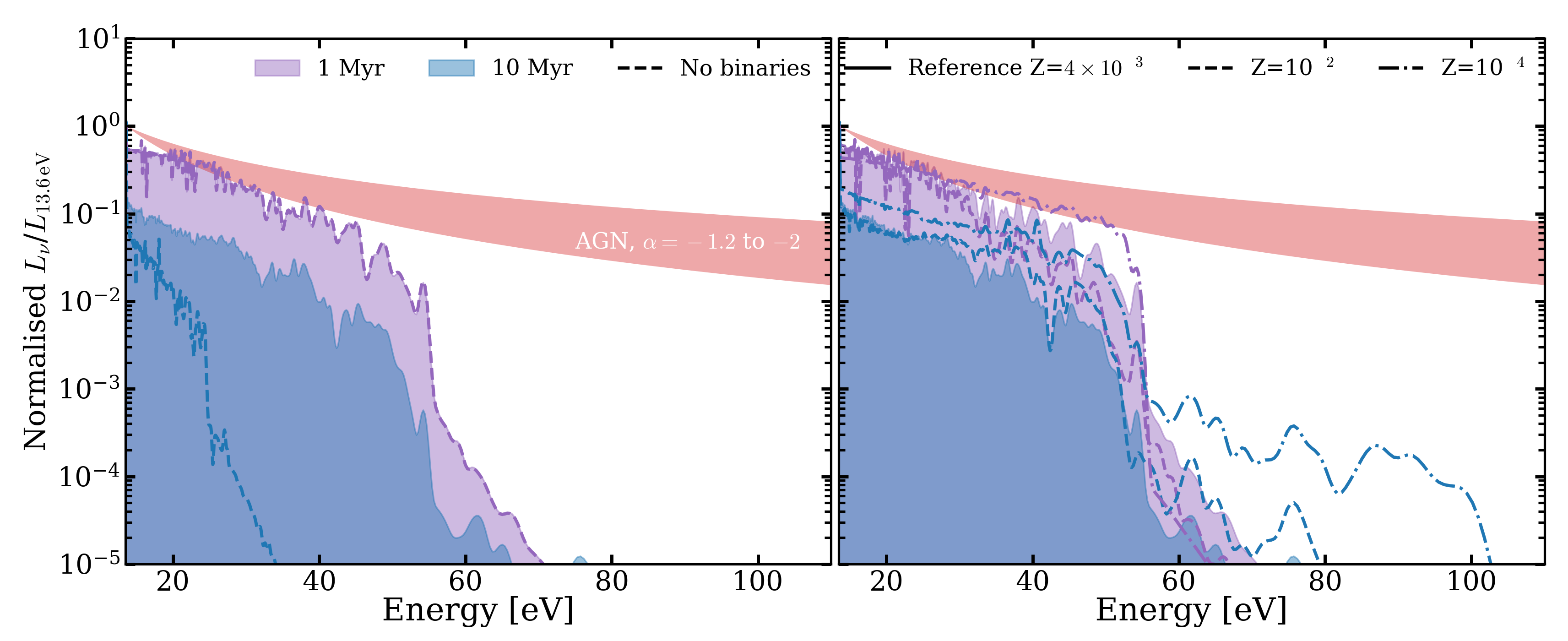} 
    \caption{The ionizing spectrum of various models of young stellar populations (blue, purple) and AGN (red) normalised to the flux density at 13.6 eV (912 {\AA}), inspired by similar figures in the literature \citep{Feltre16,Gotberg19}. In both panels, the filled purple and blue spectrum are for a reference model (Stellar population with binaries as in Fig. $\ref{fig:starsed}$, with metallicity Z=0.004 and $\alpha$/Fe=+0.6) with an age of 1 and 10 Myr. The red shaded regions shows a power-law spectrum with slope $f_{\nu}\propto\nu^\alpha$, where $\alpha=-1.2$ to $-2$ characteric for accretion disks. In the left panel, dashed lines show the spectrum without binary interactions. In the right panel, dashed lines shows a model with a higher metallicity while dashed-dotted lines show a lower metallicity model.}\label{fig:hardsed}
\end{figure*}

Due to its impact on the nebular spectrum of high-redshift galaxies, we now focus on the ionizing ($\lambda<912$ {\AA}, $>13.6$ eV) part of the spectrum specifically. Photons with energies above 13.6 eV ionize hydrogen atoms, leading to ionised regions of gas around the hottest stars. Recombinations of the hydrogen atom with free electrons lead to a cascade of transitions, causing the radiation of emission-lines, most prominently the Lyman and Balmer series. The strength of these emission-lines are sensitive to the ionizing luminosity from a stellar population, which traces the amount of very massive, short-lived stars and thus the star formation rate \cite[e.g.][]{KennicuttEvans2012,Kramarenko25}. The conversion between H$\alpha$ luminosity and star formation rate depends on numerous factors such as the fraction of ionizing photons that actually lead to recombinations (which inversely traces the escape fraction of ionizing photons), the ionizing luminosity per unit stellar mass formed and the dust attenuation.

Figure $\ref{fig:hardsed}$ shows the ionizing spectra of various models (varying the age, metallicity, and the inclusion of binary interactions) and accreting black holes, normalised to the luminosity density at 13.6 eV. The total ionizing luminosity is very sensitive to age, though binary interactions strongly increase the ionizing output at ages of 10 Myr (left panel). At fixed age, a lower metallicity increases the ionizing luminosity and slightly hardens the spectrum in the 20-50 eV range. Binary interactions at low metallicity, such as mass transfer and the development of common envelopes, also tend to make the spectrum harder above 50 eV (blue curves, 10 Myr, right panel), and they prolong the duration of significant ionizing photon emission. The relative strengths of emission lines are sensitive to the spectral hardness as various transitions have different ionization energies. For example, the [O{\sc iii}]/[O{\sc ii}] ratio is sensitive to spectral hardness as [O{\sc iii}] requires ionization energies $>35.1$ eV (whereas [O{\sc ii}] requires 13.6 eV, as H{\sc i}). Stellar spectra steeply drop above energies of 40-50 eV. This means that transitions such as C{\sc iv} and He{\sc ii} (with ionization energies as high as 47.9 and 54.4 eV, respectively) are very sensitive to the shape in this regime and therefore they are strong indicators of the presence very young and metal poor stars.

\subsection{Emission from active galactic nuclei}
In addition to stellar light, a significant -- in some cases dominant -- source of light in distant galaxies can be an active galactic nucleus (AGN). This is a luminous region in the center of a galaxy that emits light that is powered by the accretion of matter onto a supermassive black hole (SMBH). Accreting matter forms a disk that heats up and emits emission across the electromagnetic spectrum. Moreover, processes associated with highly energetic particles give rise to nonthermal emission, such as synchrotron emission in jets, which leads to bright radio and X-Ray luminosities.

AGNs are usually identified through their luminous X-Ray or radio emission, or through emission-line features in the rest-frame UV to optical spectrum that trace the hard ionizing spectra of accretion disks or the high velocities around the SMBHs. The light from AGNs varies on timescales of a few hours to months which yields key evidence for their nature as accretion disks around SMBHs. This is because the shortest variability timescale is about the light-crossing timescale. Combined with their exceptional luminosity, rapid variability thus suggests a highly powerful energy source over a very small region. Dynamical and `light-echo' techniques that measure delays in light variations across various tracers imply that the masses of SMBHs span the range $\sim10^{6-10}$ M$_{\odot}$. Such direct measurements are only available for a limited number of AGNs, but they have been used to derive calibration between black hole mass and properties as emission-line widths and luminosities \citep[e.g.][]{GreeneHo2005,Vestergaard09}

Describing the X-Ray to radio spectrum of AGN requires a whole review on its own \citep[e.g.][]{Richards06,Siebenmorgen15}. Here, we focus on the rest-frame UV and optical spectrum of AGNs. The particular emphasis of this subsection is to understand how AGN activity may mimic features of young very massive stars and, vice versa, how the presence of an AGN may be identified without usual indicators (X-Ray or radio emission), as their sensitivity is currently not very constraining in identifying faint AGN activity at $z>6$. The ionizing spectrum of an AGN is typically characterised by a power-law from the accretion disk. The accretion disk has a temperature gradient such that the combined spectrum of the blackbody emission from various regions in this disk yields a power-law like spectrum.

In the non-ionizing part of the UV spectrum (e.g. around 1500 {\AA}), the slope of an AGN spectrum is relatively similar to the spectral slope of a young stellar population. Since star-forming regions in the earliest galaxies can be extremely compact, distinguishing the contribution of AGN from star formation based on imaging data is therefore very challenging. As illustrated in Figure $\ref{fig:hardsed}$, the main difference between AGN spectra and those of young stellar populations is the lack of a strong cut-off at ionization energies of $\sim50$ eV. Therefore, transitions as N{\sc v} (77.5 eV) and [Ne{\sc v}] (97.1 eV), that require ionisation energies that are hardly emitted by stellar populations, are more distinctive AGN signatures than lines as C{\sc iv} and He{\sc ii} that could be powered by hot stars.

Recent JWST observations have revealed a population of faint AGNs that mimics spectral signatures of old stellar populations (such as a Balmer break), which makes identifying AGN contributions to the spectrum of galaxies even more challenging. These developments are discussed in Section $\ref{sec:LRDs}$.

\begin{table}[h] 
    \centering
    \small
    \begin{tabular}{cccC{7.2cm}}
    {\bf Line} & {\bf $\lambda_{\rm vacuum}$  [{\AA}]} & {\bf $\chi$ [eV]} & {\bf Example use case} \\ \hline
Ly$\alpha$ & 1215.67 & 13.6 & Redshift identification, probe of neutral hydrogen gas conditions \\
N{\sc v} & 1238.8, 1242.8 & 77.5 & Shape of ionizing spectrum \\
{[N\sc iv]}, N{\sc iv}] & 1483.32, 1486.5 & 47.4 & Ionizing spectrum, Nitrogen abundance, Electron density \\ 
C{\sc iv}  & 1548.2, 1550.8 & 47.9 & Ionizing spectrum, Carbon abundance \\
He{\sc ii}  & 1640.4 & 54.4 & Shape of ionizing spectrum \\
O{\sc iii}] & 1661.2, 1666.1 & 35.1 & Electron temperature, Oxygen abundance \\
C{\sc iii}], [C{\sc iii}] & 1906.7, 1908.7 & 24.4 & Electron density, Carbon abundance\\ 
Mg{\sc ii} & 2796.8, 2803.8 & 7.6 & Cool gas flows, SMBH mass  \\
{[O\sc ii]} & 3727.1, 3729.9 & 13.6 & Ionization parameter, electron density, Oxygen abundance \\
{[Ne\sc iii]} & 3869.7, 3970.0& 40.9 & Ionization parameter \\ 
H$\delta$ & 4102.9 & 13.6 & Ionizing luminosity, dust attenuation \\
H$\gamma$ & 4341.7 & 13.6 & Ionizing luminosity, dust attenuation  \\
{[O\sc iii]} & 4364.4 & 35.1 & Electron temperature  \\
H$\beta$ & 4862.7 & 13.6 & Ionizing luminosity, dust attenuation, Oxygen abundance, SMBH mass \\
{[O\sc iii]} & 4960.3, 5008.2 &  35.1 & Redshift identification, ionisation parameter, O abundance, ionized gas flows \\
He{\sc i} & 5877.3 & $<24.6$ & Ionisation state, Helium abundance  \\ 
{[N\sc ii]} & 6549.9, 6584.4 & 14.5 & N abundance, excitation mechanism \\
H$\alpha$ & 6564.6 & 13.6 & Ionizing luminosity, dust attenuation, Oxygen abundance, SMBH mass\\
{[S\sc ii]} & 6718.9, 6733.2 & 10.4 & Electron density \\
He{\sc i} & 7065.00 & $<24.6$ & Ionisation state, Helium abundance  \\
{[S\sc iii]} & 9071.5, 9547.6 & 23.3 & Ionisation parameter \\ \hline
    \end{tabular}
    \vspace{0.2cm}
    \caption{A list of commonly detected emission-lines from star-forming galaxies in the $\lambda_0 = 0.1-1$ micron wavelength range, their vacuum wavelength, their ionization energy $\chi$ and their typical use case.}
    \label{tab:emlines}
\end{table}

\subsection{Nebular emission}
Nebular emission is an important component in the SED of young galaxies. It predominantly originates from gas in H{\sc ii} regions (regions of ionised gas around hot stars) and can both be line and continuum emission. A detailed investigation of the various mechanisms that lead to nebular line emission (e.g. photoionization, collisional ionization) and radiative transfer processes can be found in books \cite{Osterbrock06,Draine11}, or more recent papers that describe the implementation of nebular emission in SED fitting codes \citep[e.g.][]{Li25_Cue}.

The strongest nebular lines that are typically accessible at redshifts $z\gtrsim4$ are hydrogen lines as Ly$\alpha$ and the Balmer series (e.g. H$\alpha$, H$\beta$, H$\gamma$) and collisionally excited lines from ionised oxygen ([O{\sc ii}], [O{\sc iii}]). The rest-frame wavelengths in vacuum\footnote{ Historically, air wavelengths (i.e. the observed wavelength of a photon from an emission-line after refracting in the atmosphere) have been used in the literature, for example by the SDSS survey for which the strong [O{\sc iii}] line is at 5007 {\AA} (vacuum: 5008.24 {\AA}) and H$\alpha$ is at 6563 {\AA}. For consistency with measurements from space telescopes and e.g. ALMA, vacuum wavelengths should be used, for example when using emission-lines to measure redshifts.}, required ionization energies, and typical use cases of these and fainter lines are listed in Table $\ref{tab:emlines}$. At $z>10$, measuring these lines is more difficult as current spectroscopic capabilities are less sensitive beyond 5 micron, but various UV lines have been detected out to $z\approx14$ \citep{Maiolino24,Naidu25}. Emission-line data are primarily used to i) identify and measure redshifts, ii) measure gas dynamics (to infer dynamical masses or probe outflow kinematics), iii) identify galaxies with AGN (through line-profiles and/or photo- and shock-ionization modeling), iv) measure the physical conditions of the interstellar medium (ISM), such as gas phase metallicity, metal abundances, the density of the gas, the dust attenuation, the escape fraction of ionizing photons and v) to trace the extreme UV ($\lambda<912$ {\AA}) part of the stellar spectra that cannot be directly observed due to absorption from interstellar neutral gas, but it does directly impact emission-lines in ionized regions.

\begin{figure*}[h]
\centering
    \includegraphics[width=13cm]{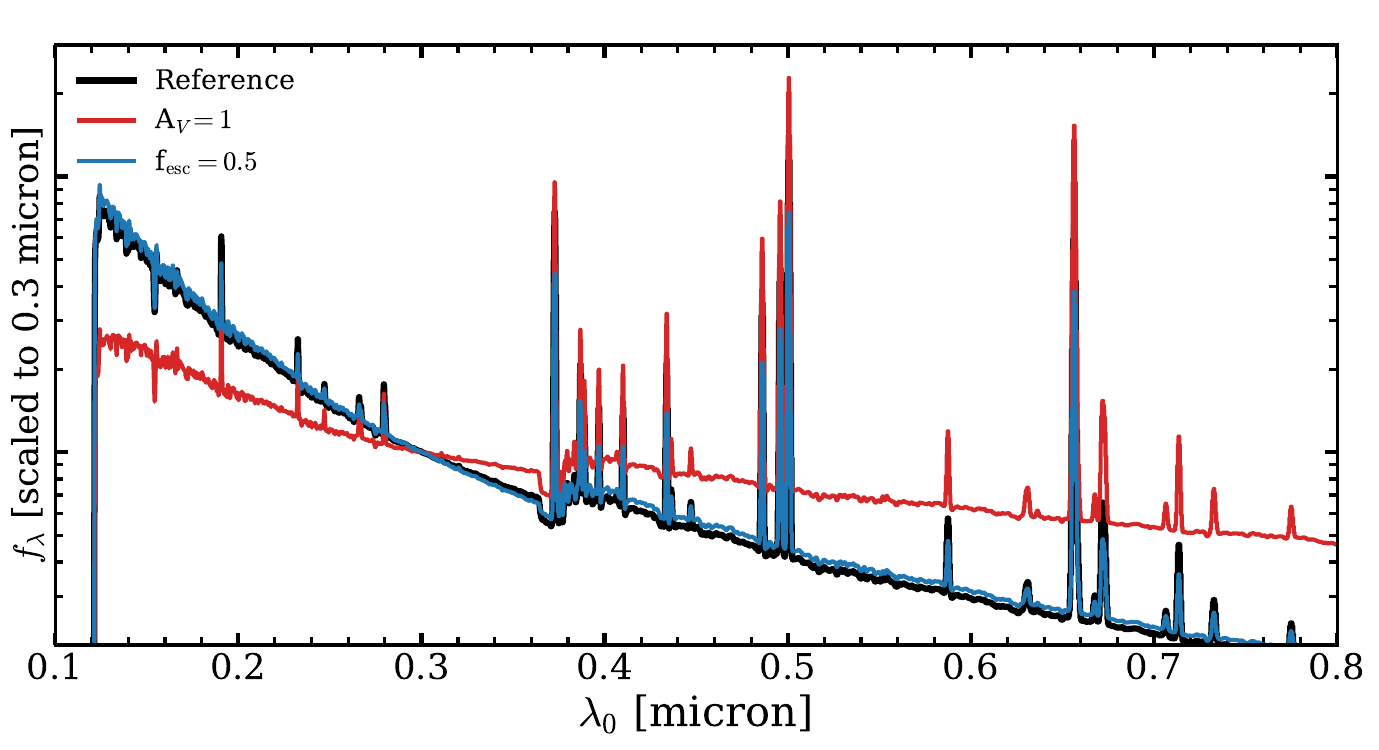}
    \caption{Rest-frame UV to optical model spectra of a young high-redshift galaxy with stellar and nebular emission. The models are constructed using BAGPIPES \cite{Carnall18} and normalised to the flux density at 3000 {\AA}. The fiducial model (black) has no dust attenuation and all ionizing photons are converted into nebular emission ($f_{\rm esc}=0$). The blue model has no dust, but $f_{\rm esc}=0.5$, whereas the red model has a dust attenuation of $A_V=1$ and a zero escape fraction. The differences in the spectra show how dust leads to reddening, whereas the effects of a high escape fraction are subtle but discernable in the continuum around the Balmer limit.  }\label{fig:dust}
\end{figure*}

Nebular continuum light can be emitted through free-bound or free-free transitions, or the two-photon process \citep[e.g.][]{Raiter10,Izotov11}. In some of the youngest galaxies that have been detected with JWST, a substantial fraction of the continuum emission has been argued to be of nebular origin, in which case Balmer jumps (i.e. a depression in the continuum level rightward of the Balmer break) can be detected \citep{Cameron24,Katz24}. The detailed shape of the nebular UV continuum is sensitive to the gas conditions such as the electron density and temperature. The presence of nebular continuum emission in young galaxies is important to include in SED modeling as it tends to redden the UV continuum (which could otherwise be misinterpreted as dust attenuation), unless significant numbers of ionizing photons escape the ISM, in which case the strength of nebular emission decreases. These subtle effects can be seen by comparing the reference model with the model with ionizing photon escape in Figure $\ref{fig:dust}$.

\subsection{Dust attenuation and emission}
One of the key contents of galaxies is interstellar dust. Dust particles may scatter and absorb light, and the absorbed energy is re-radiated in the (far) infrared. Accounting correctly for the dust attenuation\footnote{While {\it extinction} accounts for absorption and scattering of light outside of the line of sight, {\it attenuation} additionally also accounts for differences in dust column densities to different background sources and scattering into the line of sight.} of light due to dust grains is critical to recover the intrinsic stellar, nebular and/or AGN emission. The luminosity and shape of the infrared dust emission spectrum, which typically follows a grey body shape with peak temperatures ranging from $\sim20-100$ K, are sensitive to the amount of energy absorbed, and hence the ionizing output from a galaxy and the properties and distribution of dust. 

The attenuation curve depends on the physical properties of the dust (such as the sizes and compositions of the grains) and geometry and it is common practice to assume locally calibrated attenuation curves, such as the Calzetti curve \citep{Calzetti99} or the dust attenuation curve derived in the Small Magellanic cloud \citep{Gordon03}. The attenuation for stellar light is often different than the attenuation of the nebular emission \citep[e.g.][]{Reddy15}, which also seems to follow a different attenuation law \citep{Cardelli89}. Most analyses assume relatively simple geometries, such as a single screen of dust in between the observer and the stellar population. Dust attenuation can significantly impact the inferred properties from galaxies, such as the star formation rate or the stellar mass. For these reasons, understanding which attenuation curve and geometry to use in which situation is a topic of ongoing research.

Various methods have been developed to accurately account for dust attenuation. They predominantly use the fact that dust attenuation reddens the observed spectrum, meaning that relatively more light is attenuated at shorter wavelengths (see Figure $\ref{fig:dust}$). For example, the reddening can be inferred by comparing the measured UV slope to an assumed intrinsic UV slope of a stellar population \citep{Meurer99}. In SED fitting codes the reddening is included as a free parameter and the uncertainties due to degeneracies between variations in intrinsic colors and dust reddening are propagated. In some codes energy balance is assumed, meaning that there is a coupling between the attenuated UV light and the infrared emission \citep{Cunha08}. The attenuation of the nebular light can be estimated with the Balmer decrement\footnote{In principle this method could be extended to other lines, in particular combinations of Paschen and Balmer lines that trace the same energy levels \citep{Reddy25}.}. Under standard assumptions about the physical conditions of the gas, the intrinsic line-ratio of Balmer lines are known. Deviations from these intrinsic line-ratios are assumed to be due to differences in the attenuation at the respective wavelength of various transitions, and therefore can be used to estimate the total attenuation.

\subsection{The intervening medium in absorption}
The light from distant galaxies is also impacted by gas along the line-of-sight from the galaxy to us. The gas can be part of the interstellar medium, which is the material that resides among the stars within galaxies and that consists of atoms, molecules and dust, but in many cases the gas is part of the so-called circum-galactic medium. The circum-galactic medium is a diffuse envelope of gas that extends about 10 times the size of the galaxy, out to the virial radius, and that mostly has a lower density and higher temperature than the interstellar gas. Gas that is located outside the virial radii of galaxies is called the inter-galactic medium, which is usually even less dense.

\begin{figure}[h]
    \centering
    \includegraphics[width=0.98\linewidth]{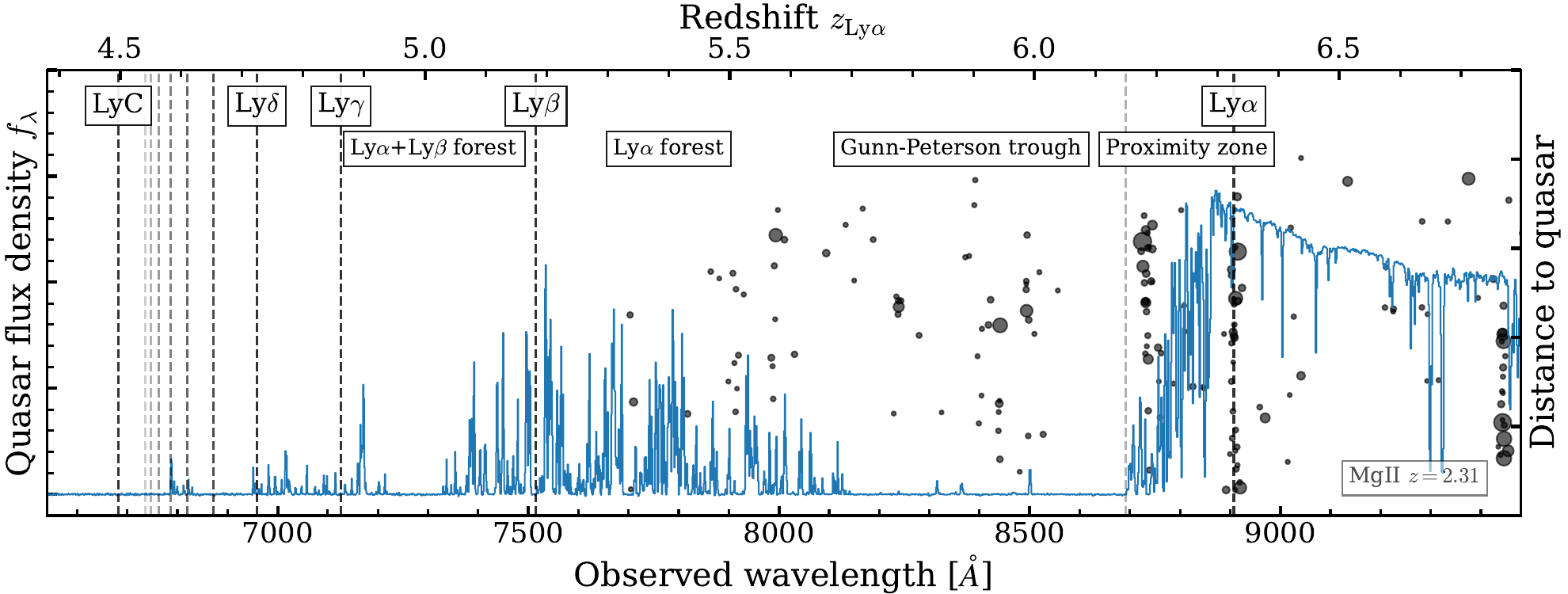}
    \caption{The rest-frame UV spectrum of the luminous quasar J0100+2802 at $z=6.327$ taken with the X-SHOOTER spectrograph on the Very Large Telescope. The spectrum highlights the high-redshift Lyman-$\alpha$ forest transmission at the end stages of reionization ($z=5.3-6.0$) a strong Gunn-Peterson trough at $z=6.0-6.15$ suggesting a neutral region and the ionized region influenced by the quasar at $z=6.15-6.30$. Absorption lines in the quasar spectrum redward of the Lyman-$\alpha$ line are due to metal transitions at lower redshifts, such as Mg{\sc ii} at $z=2.31$. Also highlighted are the positions of galaxies in the $z=5.3-6.9$ range identified with JWST/NIRCam \citep{Kashino23}, where their y-values are the line of sight distance to the quasar and the symbol size scales with luminosity.}
    \label{fig:quasar}
\end{figure}

Diffuse gas can be detected through relatively sharp absorption lines to the background sources and it is particularly sensitive to phases of gas that are otherwise not very luminous, such as diffuse neutral hydrogen. The absorption line strength, velocity and width can reveal physical insights such as the temperature, ionization state and abundances of the gas. As such features require a high signal-to-noise spectrum, they are typically detected in luminous objects such as quasars (see Figure $\ref{fig:quasar}$ for an example quasar spectrum that shows various of the discussed types of absorption).

The most prominent types of absorption feature are: i) neutral and ionised gas in the interstellar medium from the galaxy, which is usually identified with metal transitions and is used to study the presence of outflows and the covering factor \citep{Gazagnes24,Snapp25}; ii) relatively dense neutral hydrogen clouds that cause so-called Damped Lyman-$\alpha$ Absorptions that can be identified in the ISM of the emitting galaxy, but also frequently in the foregrounds of quasar and galaxy spectra; iii) metal absorption lines of enriched gas in the circum-galactic medium around galaxies that indicate significant amounts of enriched gas exists around galaxies in a range of temperature-density states \citep[e.g.][]{Bordoloi24}, iv) finally, relatively diffuse (neutral) hydrogen gas that traces relatively low over-densities of gas between galaxies visible as the Ly$\alpha$ forest. It is in the diffuse gas between stars and around galaxies where the interactions between physical processes associated to stars (star formation, supernova explosions) and the large-scale structure (gas inflows, galaxy mergers) happen. Therefore, the state of this gas is an important testbed for models of galaxy formation.

As we approach the epoch of reionization, the neutral fraction of the gas between galaxies (the inter-galactic medium; IGM) increases, leading to saturated Ly$\alpha$ absorption at $z\approx6$ (at neutral fractions $\sim10^{-5}$, visible as Gunn Peterson troughs; see Figure $\ref{fig:quasar}$, \cite{FBS23}). The relation between the intergalactic Ly$\alpha$ transmission and the distance to galaxies is being investigated around redshifts $z\sim6$ \citep{Kashino23,Kakiichi25} as a probe of the role of galaxies in the reionization of the Universe, respectively, see Section $\ref{sec:reion}$.

\section{High-redshift galaxy selection techniques} \label{sec:find}
\begin{figure*}
    \centering
    \includegraphics[width=15cm]{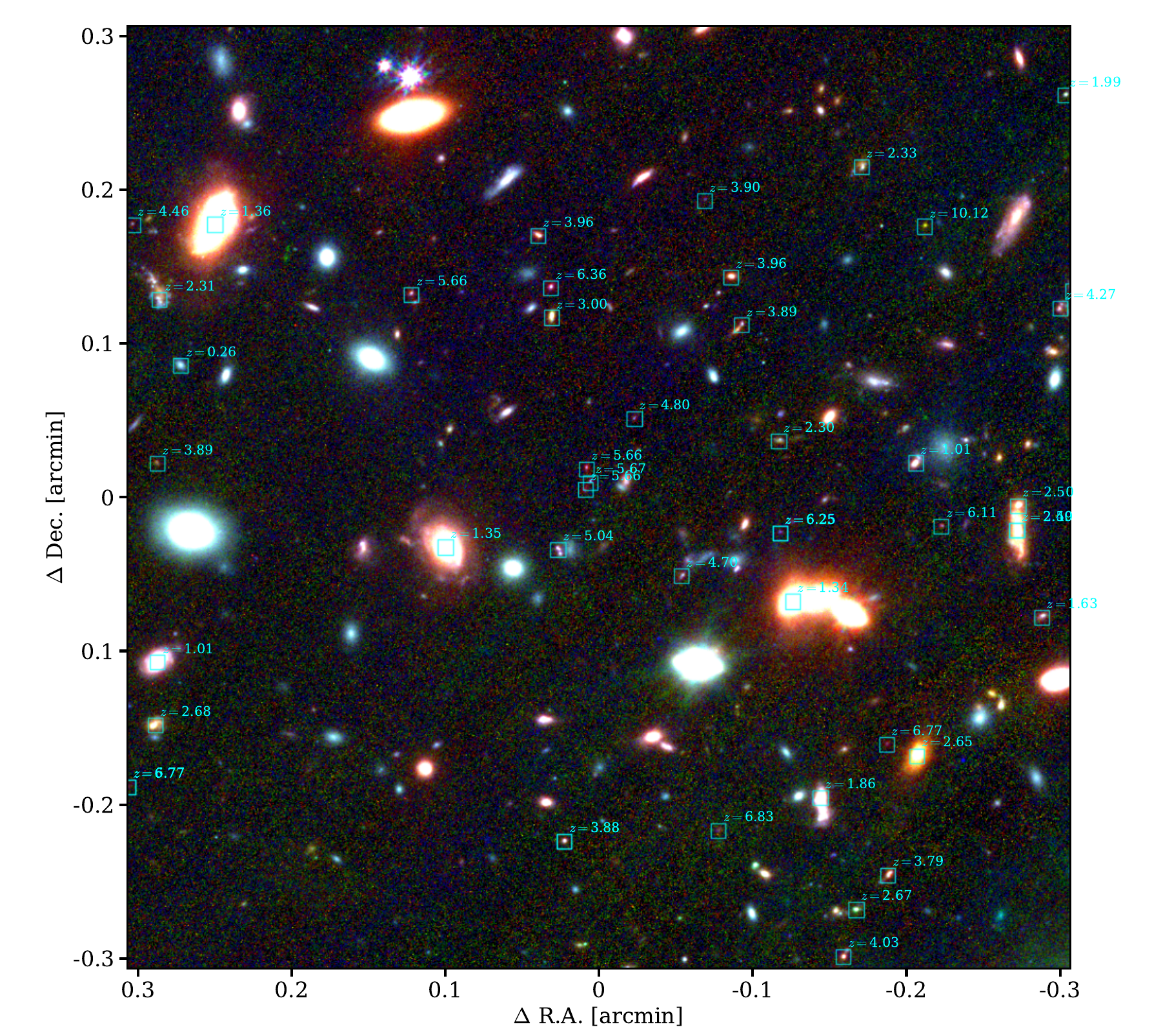} \\
    \caption{An arbitrary chosen $0.6\times0.6$ arcmin$^2$ region in the larger field around the Abell 2744 lensing cluster observed by JWST \citep{Naidu24,Bezanson24}. The false-color image is constructed with the F356W/F200W/F115W filters on NIRCam, respectively. Cyan squares highlight galaxies for which JWST has measured their redshift spectroscopically \citep{BWang23}, spanning $z\approx1-10$ (i.e. light-travel times of $\sim$7 to 13.2 billion years) and illustrating the variety of galaxy colors and sizes that are indicators of their redshifts.} 
    \label{fig:ALT_view}
\end{figure*}

The most distant galaxies in deep sky images are usually not larger than a few pixels and they are vastly outnumbered by other sources of light, such as faint stars in the Milky Way and less distant galaxies. Various techniques have been developed to detect and identify galaxies in the early Universe efficiently. A key underlying aspect of these techniques is that the range of possible spectra of galaxies is relatively constrained and well understood (see Section $\ref{sec:spectrum}$). Moreover, sensitive instruments on the most powerful telescopes have the capabilities designed to identify specific spectral features that can pinpoint the redshift (and hence the cosmological distance) to galaxies. The state of the art is illustrated in Figure $\ref{fig:ALT_view}$, which shows a deep JWST/NIRCam image with a large number of redshifts that are measured spectroscopically. The light of virtually every galaxy on this image was emitted before the Sun formed and the faintest sources of light reach 1 nanoJansky (i.e. about a million times fainter than a single candle at the distance of the moon) and were emitted when the Universe was less than 500 million years old. 

\subsection{Observational facilities and strategies}
Our ability to find the most distant galaxies depends primarily on the photon-collecting power of telescopes that scales with mirror area. However, other important factors that determine our sensitivity to distant objects are the background level, the angular resolution and the spectrum of distant sources. The main telescopes that are being used in the study of the most distant galaxies are optical telescopes that observe in the $\lambda\approx0.1-10$ micron (i.e. UV-Optical-Infrared) range, such as the Very Large Telescope, the HST and the JWST. Important and unique contributions, in particular of studies of the gas and dust in distant galaxies, are also being made by the Atacama Large Millimeter Array (ALMA). 

For UV to near-infrared wavelengths, electronic charge-coupled devices enable a stable record-keeping of the number of observed photons from a given patch of sky. This makes it possible to combine exposures taken over a range of dates to maximise the signal to noise ratio. The main sources of noise on these detectors are due to electric currents generated during the read-out of the detector and photons from foreground or background sources. These originate from the glow of our own atmosphere and/or light from gas in the solar system or the Milky Way. Another important consideration is the limited resolution of telescopes. While the diffraction limit ($\approx1.22'' \lambda/D$, where $\lambda$ is the wavelength and $D$ the diameter of the aperture) of a telescope, as the Very Large Telescope at 0.5 micron is around 0.01$''$, the typical resolution reached on visible images is about a factor 50 lower due to atmospheric turbulence yielding a typical image resolution of $\sim1''$. Correction techniques, such as (laser guided) adaptive optics, typically manage to improve this resolution close to the diffraction limit.

Currently, the most sensitive ground-based telescopes are located on sites with excellent atmospheric conditions (with a seeing of around 0.5$''$), a dark sky with low light pollution, and at high altitudes to reduce the likelihood of cloud coverage and to reduce the atmospheric water vapor column (the latter particularly important for sub-mm observatories such as ALMA). These sites are usually high mountains close to large oceans that provide a stable atmosphere, such as Maunakea on Hawaii, the Andes mountains in Chile and the Canary Islands in the Atlantic. 

There are various reasons to launch telescopes to space, in particular to not suffer from atmospheric turbulence and to overcome the transmission and background emission limitations from the Earths atmosphere. These are particularly important at short wavelengths ($\gamma$ rays, X-Rays, UV) that the atmosphere blocks and infrared wavelengths that the atmosphere emits itself. The typical resolution that the 2.5m Hubble Space Telescope reaches in the optical is about 0.05$''$, a vast improvement over ground-based observatories. While the deepest near-infrared images from ground-based observatories reach sensitivities $\sim100$ nanoJansky (nJy) \citep{Brammer16}, the space-based JWST NIRCam images reach sensitivities of $\sim1$ nJy \cite{Eisenstein23,Kokorev25}. Another benefit of space-based observatories is the amount of available dark hours per day (almost 24 hours for telescopes at L2 such as JWST, versus about eight for a typical ground-based observatory). The downsides of space telescopes are the significant higher costs, the inability of fixes or upgrades to the telescope (in case space-telescopes are in solar orbits like L2) and weight and size limitations.

The majority of studies of the distant Universe are confined to several well-selected survey fields. In order to optimise the detection of faint sources these are usually selected to be mostly devoid of Milky Way stars or large nearby galaxies. However, other considerations are observability, such as being accessible by telescopes on both hemispheres (the well-known COSMOS field is such an equatorial field), year round observability (near the poles), the availability of legacy data over several years and wavelength ranges, or the presence of peculiar objects such as powerful lensing clusters or luminous quasars.

\subsection{High-redshift galaxy selections before JWST} \label{sec:prejwst_search} 
Finding the most distant galaxies in images that contain over 10,000s of sources requires sophisticated and (highly) automated selection techniques. Multiple such techniques have been developed to search for various specific strong spectral features in the spectra of galaxies. These techniques are roughly split into two classes: continuum selections and emission-line selections. Continuum selections typically use photometry in broad-band filters, while emission-line selections rely on narrow-band photometry or spectroscopic data. From the 1990s-2020s, the most commonly used techniques to find galaxies in the first 3 Gyrs ($z>2$) relied on two of the strongest rest-frame UV features that are redshifted in the favorable optical wavelengths at high-redshift\footnote{Sub-mm data has also been used to identify galaxies on their strong infrared continuum emission. However, measuring their redshift has typically been challenging due to the need to perform sensitive spectral scanning in the infrared over a large bandwidth.}: the Lyman Break and the Lyman-$\alpha$ emission.  

\begin{figure}
    \centering
    \includegraphics[width=0.98\linewidth]{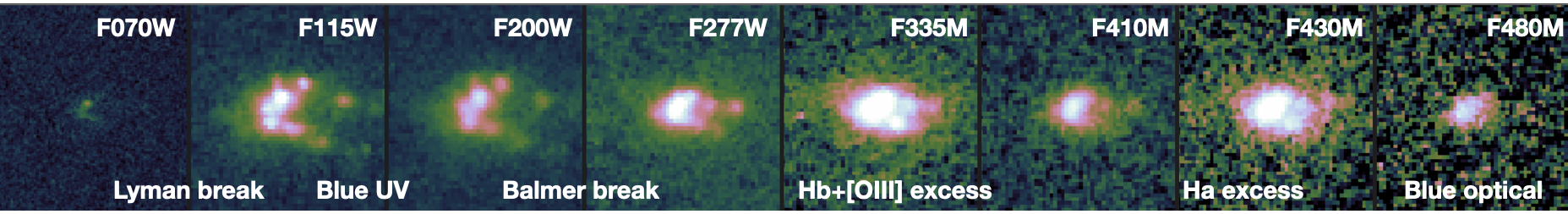}
    \caption{Zoomed-in JWST NIRCam images of a low mass ($\sim10^8$ M$_{\odot}$) galaxy at a spectroscopically measured redshift $z=5.66$, corresponding to a look back time of 12.7 billion years. The galaxy has been magnified with a magnification factor $\mu\approx20$ by the Abell 2744 lensing cluster. The images in various filters display the strong Lyman break that is clearly visible between the F070W and F115W filter, the blue UV and optical continua (from the soft fading in F115W to F200W and F410M to F480M, respectively) and the strong H$\beta$+[O{\sc iii}] and H$\alpha$ emission (in the F335M and F430M medium-bands.}
    \label{fig:stamps}
\end{figure}

The Lyman break technique, first used successfully for large galaxy samples at $z\sim3$ by Steidel et al. in 1996 \citep{Steidel96}, but already discussed at least two decades earlier \citep[e.g.][]{PartridgePeebles67,Sandage73,Koo85}, relies on strong discontinuities in galaxy spectra around the Lyman-$\alpha$ line at $1216$ {\AA} and around the Lyman-Continuum limit (912 {\AA}), see Figure $\ref{fig:sed}$. These discontinuities originate from the partial ionisation state of stellar atmospheres. At high-redshift, in particular $z>3$, neutral gas in the intergalactic medium additionally absorbs Lyman-Continuum and Ly$\alpha$ photons \citep{Madau95,Inoue14}. This effectively leads to a single (even stronger) discontinuity at the Lyman-$\alpha$ wavelength for galaxies and quasars at redshifts beyond $z>6$ due to the Gunn-Peterson trough (see for example the quasar spectrum in Figure $\ref{fig:quasar}$). The Gunn-Peterson trough is an extreme (saturated) case of similar Lyman-$\alpha$ absorption in the regime where most gas in the intergalactic medium is neutral, which was the case before the end of cosmic reionization (see also Scheuer 1965, \citep{Scheuer65}). This causes high-redshift galaxies to `drop-out' in images taken in filters that cover rest-frame wavelengths below 1216 {\AA} (see for example Figure $\ref{fig:stamps}$), which happens in optical wavelengths at $z\sim3-7$ and at higher redshifts in the near infrared. Such galaxy samples (Lyman Break Galaxies) at specific redshift ranges are usually defined by a combination of a single or multiple colour criteria, in addition to a (continuum) magnitude limit, or by evaluating SED templates at various redshifts in a likelihood analysis. Deep optical and near-infrared imaging surveys enabled the construction of large samples at $z\sim2-9$ \citep[e.g.][]{Bouwens15,Harikane22}. The increase in 1-5 micron sensitivity enabled by JWST has enabled the Lyman break technique to successfully identify galaxies out to redshifts $z\sim15$ \citep{Carniani24,Naidu25}. The main classes of interlopers to this technique are objects that show similarly strong colours, such as strongly dust-obscured and/or passive galaxies with a Balmer break (out to $z\sim5$ in $z>10$ samples with JWST; \cite{Naidu22,ArrabalHaro23}, and brown dwarf stars \citep[e.g.][]{Bowler15,GRB24}). These interlopers are particularly an issue for luminous high-redshift galaxies that are often identified in fields with shallower data or sparser wavelength coverage, and less of an issue for fainter systems as the number counts of high-redshift galaxies are typically much steeper than those of interlopers, such that the likelihood of encountering interlopers decreases towards fainter luminosities.

The most commonly used emission-line selection identifies galaxies based on their Lyman-$\alpha$ emission (hence Lyman-$\alpha$ emitters) and also dates back to the 90s \citep[e.g.][]{Hu96}. The Lyman-$\alpha$ line is intrinsically one of the brightest emission-lines produced in H{\sc ii} regions, and its rest-frame UV wavelength makes it very suitable for identifying galaxies at redshifts $z\sim2-7$. Indeed, the narrow-band technique, where emission-line galaxies are identified based on a colour excess over the photometry in an overlapping broad-band has yielded samples of thousands of Lyman-$\alpha$ selected galaxies at $z\sim2-7$ \citep[e.g.][]{Sobral18,Ouchi20}. The main contaminants for these samples are other emission-lines from lower redshift galaxies, primarily [O{\sc iii}] and H$\alpha$. Additional photometric information is used to identify such interlopers \citep[e.g.][]{Matthee17}. With the advent of wide-field integral field spectroscopy, in particular thanks to the Multi Unit Spectroscopic Explorer on the Very Large Telescope, direct spectroscopic high-redshift Lyman-$\alpha$ selection has been very efficient \citep{Bacon17}. Due to the high spectral slicing ($\Delta \lambda=1.25$ {\AA}), the contrast of narrow emission lines over the sky background is very high, yielding unprecedented (surface) brightness limits, enabling the detection of diffuse light \citep{Wisotzki18} and galaxies that are undetected even in the deepest HST data \citep{Maseda18}. 

While any flux-limited galaxy sample is an incomplete sample of the full galaxy population, an emission-line selected sample is even more incomplete, as emission-lines are only emitted in the youngest galaxies. In particular, Lyman-$\alpha$ emission-line samples represent only a subset of the galaxy population, as the radiative transfer of the Ly$\alpha$ between the production sites (H{\sc ii} regions) and telescopes is affected by dust attenuation (the UV continuum alike), but also by scattering on neutral hydrogen in the interstellar medium and the gas outside galaxies \citep[e.g.][]{Neufeld91,Gronke15}.
These processes increase the effective attenuation due to the increasing likelihood of encountering a dust particle, and diffuse the photons in the spatial and spectral direction, lowering the contrast to the background. Spectroscopic follow-up studies of continuum-selected galaxies indicate that the overlap with Ly$\alpha$-selected samples is only $\approx5$ \% at $z\sim2$, but it increases to $\approx50$ \% at $z\sim5$ \citep[e.g.][]{Stark10,Kusakabe20}, before declining again at higher redshift due to reduced transmission in an increasingly neutral intergalactic medium (see Section $\ref{sec:reion}$, \cite{Kageura25}). The increase is likely a result of the decrease in age and dust attenuation in typical galaxies from $z\sim2$ to $z\sim5$ \citep{Hayes11}.

Before JWST, directly selecting galaxies with rest-frame optical emission-lines at high-redshift has been challenging as H$\alpha$ and [O{\sc iii}] could only be selected out to $z\approx2,3$, respectively. Yet, samples at these redshifts have been created with near-infrared narrow-band observations in the first two decades of this millennium \citep[e.g.][]{Khostovan15,Terao22}. Additionally, despite limitations in its sensitivity, spatial resolution and the use of very wide filters, photometry with the {\it Spitzer} space telescope has been used to identify galaxies with strong lines at $z\sim5-9$ \citep[e.g.][]{GRB16,Strait21}.

\subsection{High-redshift galaxy selections in the JWST era} \label{sec:highz_jwstera}
Since the summer of 2022, the JWST has made immense improvements in our ability to identify and study distant galaxies. This is enabled by various key technological advances thanks to JWST's large passively cooled 6m mirror and the diverse instrument suite. The Near-InfraRed Camera (NIRCam) enables sensitive imaging in various filters covering the $\sim0.7-5 \mu$m regime at $\sim0.05-0.1''$ resolution and wide-field slitless spectroscopy over 3-5 $\mu$m. The Mid-Infrared Instrument (MIRI) is optimised for imaging and spectroscopy over 5-30 $\mu$m at $\sim0.2-0.8''$ resolution. The Near Infrared Spectrograph (NIRSpec) enables single and multi-object spectroscopy over the 1-5 $\mu$m range, as well as an integral field mode, with a spectral resolution from 100 - 3000 km s$^{-1}$. Moreover, the Near Infrared Imager and Slitless Spectrograph (NIRISS) enables highly efficient wide-field slitless spectroscopy over $1-2 \mu$m  and imaging from 1-5 $\mu$m. The field of view of JWST's instrument ranges from 9 arcsec$^2$ (NIRspec and MIRI integral field spectroscopy) to $\sim10$ arcmin$^2$ (NIRCam imaging). Among these cameras that all significantly improve upon previous available instruments, the most significant gain has been made in the spectroscopic capabilities over the 2-5 $\mu$m wavelength regime, where the sensitivity improvement has been more than three orders of magnitude, in addition to the new multiplexing capabilities.

\begin{figure*}
    \centering
    \begin{tabular}{cc}
    \includegraphics[width=6.7cm]{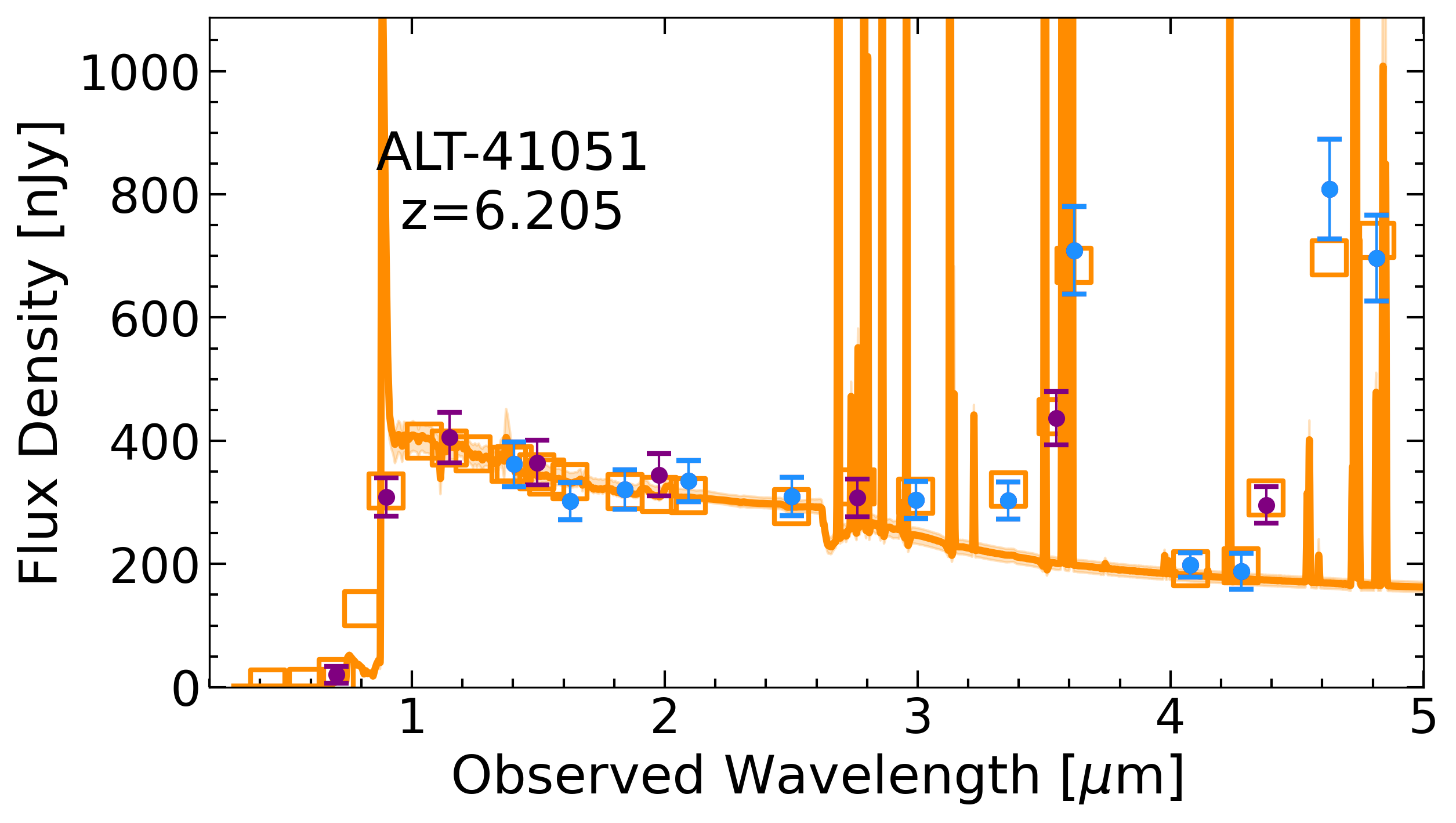} & 
    \includegraphics[width=6.7cm]{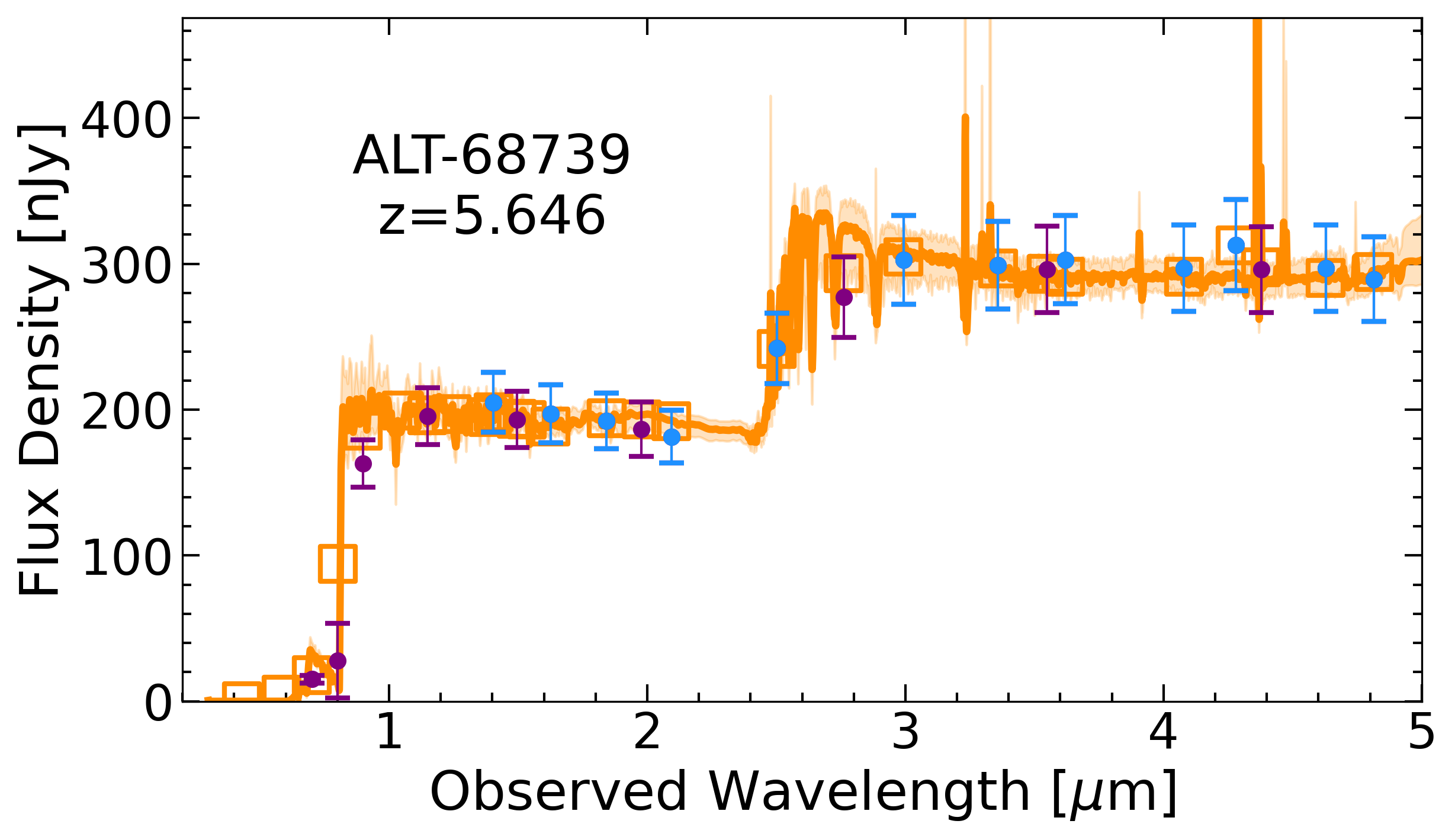} \\
    \end{tabular}
    \caption{Two example SEDs from galaxies at $z\sim6$ discovered with JWST NIRCam slitless spectroscopy from the ALT survey. We focus on $z\sim6$ as this is the redshift where JWST can detect various features ranging from the Lyman break to the H$\alpha$ line. Purple data-points show photometry in broad-band filters and blue data-points show medium-band filters. The orange curve show best-fit SEDs, and the open squares show the predicted flux density in various pass-bands. Since the redshift was measured spectroscopically, data around the Lyman break were not included in the fitting. The left galaxy shows a typical young galaxy with a blue UV continuum slope, relatively little dust attenuation and strong emission-lines. The right galaxy shows a galaxy with a significantly older stellar population, displaying a Balmer break and only weak emission-lines. } 
    \label{fig:ALT_galaxy}
\end{figure*}

The wider and redder wavelength coverage compared to e.g. HST has enabled the Lyman-break method to be used to identify candidate galaxies out to $z\approx20$ \citep[e.g.][]{Castellano25,PerezGonzalez25}. However, multi-band photometry, in particular medium-band photometry, has also enabled new rest-frame optical emission-line selections where strong lines as H$\alpha$ and [O{\sc iii}] with typical equivalent widths of $\sim500-1000$ {\AA} at $z\sim6$ \citep{Endsley24} can boost the photometry in specific bands (out to $z\sim9$; e.g. \cite{Rinaldi23,Withers23,Duncan24,Wold25}). Fine-sampled photometry enabled by a plethora of medium-band filers also enables new selections of galaxies selected specifically to have weak emission-lines and/or a strong Balmer break suggestive of an older stellar population \citep[e.g.][]{Looser24,Trussler25}. 

Figure $\ref{fig:ALT_galaxy}$ shows two example galaxy discovered by JWST. Their redshifts are determined spectroscopically at $z=6.205$ and $z=5.646$ with the NIRCam grism mode \citep{Naidu24}. The spectrum in the left panel shows a Lyman-break, with a very strong break from photometry in the F070W to the F090W and the F115W filters around 1 micron. Strong emission-lines in the rest-frame optical boost the flux density in broad-band filters by a factor $\sim2$ and in medium-bands by a factor $\sim3.5$ (see also Figure $\ref{fig:stamps}$). The SED has a blue UV continuum slope typical for relatively young galaxies with little dust attenuation. In the right panel, the SED of a galaxy with a more evolved stellar population is shown. This SED displays a Balmer break (in addition to the Lyman break), a redder UV continuum and there are relatively weak emission-lines. 

The wide field slitless spectroscopic mode of NIRCam has been particularly effective at producing large galaxy samples of emission-line selected sources with spectroscopic redshifts at $z>3$ \citep{Kashino23,Oesch23,Naidu24,Lin25,Sun25Sapph}. This mode benefits from the relatively low background around 4 micron and the relatively high spectral resolution for a grism ($R=\Delta \lambda/\lambda \approx 1600$), which enables the separation of narrow emission-lines from continuum contamination from foreground objects, as well as the redshift identification of galaxies thanks to the bright [O{\sc iii}] doublet. The identification of H$\alpha$ lines usually requires photometric data to distinguish the line from other single emission-lines (such as Paschen-$\alpha$ or Paschen-$\beta$ from galaxies at $z\sim1-2$).

To date, photometric samples in deep extra-galactic fields have yielded samples of $\sim1,000,000$ galaxies with photometrically identified redshifts out to $z\sim15$ \citep{Eisenstein23,Merlin24,Weibel24,Suess24,Finkelstein25,Shuntov25,Sarrouh25}. The number of spectroscopic redshifts, determined primarily from follow-up observations with NIRspec but also identified with NIRCam slitless spectroscopy, is on the order of $\sim10,000$ \citep[e.g.][]{Naidu24,Sun25Sapph,dEugenio25jades,Mascia24,Price25,deGraaff25}\footnote{These numbers are rough estimates in Summer of 2025 and are rapidly increasing.}. Rest-frame optical emission-line measurements are efficient up to redshifts $z\sim10$ (above which [O{\sc iii}] redshifts beyond 5.5 micron and is no longer observable with NIRspec). Lyman break selections are possible out to the highest redshifts at which galaxies exist, but above $z>10$ spectroscopic confirmation of the redshift primarily relies on detecting the continuum break (although strong emission-lines are notable and have been detected; see Section $\ref{sec:earlySF}$). MIRI spectroscopy is significantly more expensive than spectroscopy with NIRspec, primarily because of the lack of a multiplexing option, but also due to a sensitivity drop. Nevertheless, rest-frame optical emission lines have been directly detected at $z>10$ \cite{Calabro24,Alvarez25}, and are also seen contaminating broad-band photometry \citep{Helton25}, facilitating redshift identification through modeling of photometry (so-called photometric redshifts). There is usually significant overlap among galaxy samples that are identified through JWST photometry and spectroscopy as the typical emission-line equivalent widths of the rest-frame optical lines are very high \citep{Boyett24}. As discussed in Section $\ref{sec:bursty}$, at least some relatively young galaxies are being identified at high-redshift without strong emission-lines, indicating that samples are not fully overlapping.

\section{Recent developments, open questions and controversies} \label{sec:science} 

In this final part of the review, we aim to provide an overview of research topics that have been particularly actively studied, emphasizing the main objectives and new methods and datasets that are driving progress and what are currently some of the main limitations. This list of topics is certainly incomplete and biased by the author's personal work and background, and barely touches upon areas such as studies of galaxy structures, kinematics, spatially resolved properties or the properties of dust (see \cite{Adamo24,Stark25,Ellis25} for other overviews of recent results on these topics).

\subsection{Early star formation} 
The history of star formation in the Universe is among the most fundamental properties that observational astronomers aim to infer. One of the main advances of the {\it JWST} is that it enables us to obtain more and better probes of galaxies' star formation activity beyond redshifts $z>2$. Before {\it JWST}, the most sensitive constraints on the evolution of the cosmic star formation rate density were primarily based on rest-frame UV data, taken, for example, with the {\it Hubble Space Telescope} \citep[e.g.][]{Bouwens15}. These measurements were affected by two main limitations. First, the observed UV light from young stars may be strongly attenuated by dust in the interstellar medium of galaxies up to levels that samples of strongly obscured galaxies may have been missing from the samples. Direct measurements of the obscured star formation rate density have been made based on infrared and sub-millimeter data \citep{Gruppioni13,HodgeCunha20}, but these have typically not reached sensitivities comparable to rest-frame UV data. Second, the redshifts of the vast majority of galaxy samples at $z>2$ were estimated based on photometry, which may contain biases or be subject to interlopers. With {\it JWST} these challenges can be addressed thanks to the large number of spectroscopic redshifts and the possibility to select galaxies by their H$\alpha$ line emission (a sensitive tracer of ongoing star formation that is significantly less subject to dust attenuation; \citep{Sun25,diCesare25}). Moreover, {\it JWST} data have advanced the redshift frontier ($z\sim10-15$), yielded new insights into the star formation histories of individual reionization-era galaxies, and probed the assembly of the first massive galaxies.

\begin{figure}[h]
    \includegraphics[width=14cm]{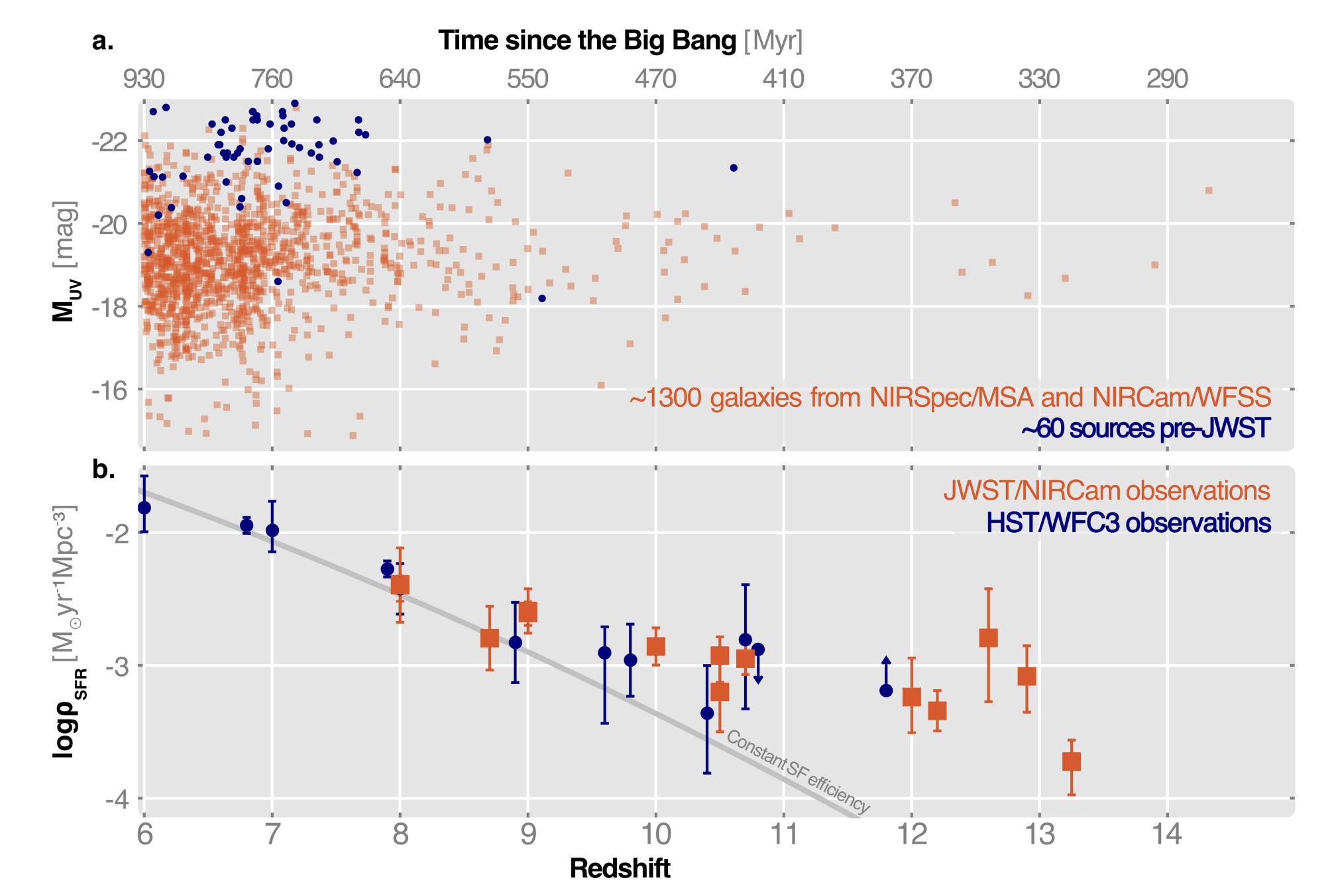}
    \caption{The advances of probing the first Gyr ($z\gtrsim6$) made by JWST. Top: The spectroscopically measured redshifts and UV luminosities of galaxies in the early Universe based on JWST (orange) and pre-{\it JWST} (blue) data. Bottom: the evolution of cosmic star formation rate density in the first Gyr based on UV data. New JWST results beyond $z>10$ suggest an unusually efficient production of UV light exceeding expectations from pre-JWST models (grey line). Figure adapted from \citep{Adamo24}.} \label{fig:UVLF}
\end{figure}

\subsubsection{A bright cosmic dawn} \label{sec:earlySF}
The number of (luminous) galaxies discovered by {\it JWST} at redshifts beyond $z>10$ have defied expectations \citep{Naidu22,Castellano22,Adams23}. Although the data-sets that reach into new regions of parameter space have also suffered from new classes of interlopers (such as dust-obscured emission-line galaxies at redshifts $z\sim5$, \citep{Naidu22,ArrabalHaro23}), multiple studies reported a factor of a few more luminous galaxies at $z>10$ \citep{Bouwens23,Donnan24,Whitler25}, compared to the expectations (see Figure $\ref{fig:UVLF}$). These expectations were based on models that assume a constant efficiency of star formation from gas accreting on dark matter halos, whose evolution can be predicted for a given cosmological model, and a fixed efficiency of UV light production per unit star formation rate \citep{Mason15,Tacchella18}. These models successfully matched the evolution of the UV luminosity density after the first Gyr of cosmic time. What could be causing these unexpectedly high numbers of early galaxies? While there are models that accelerate structure formation itself (effectively changing our cosmological model, see \cite{MKB23} for a discussion), the majority of possible explanations are astrophysical. One suggestion is that the star formation efficiency was higher in the early Universe, for example, because supernova explosions may not strongly suppress star formation (so-called supernova feedback) in early galaxies. This may happen in the high gas densities in early galaxies, where the supernova feedback is delayed with respect to the timescales on which stars form \citep[e.g.][]{Dekel23}. Alternatively, a more efficient conversion of stellar mass to UV light could yield a higher number of galaxies without the need to form more stellar mass. This could be achieved either by a negligible dust attenuation \citep{Ferrara23} or by processes that change the relative fraction of massive stars that contribute to the galaxies' spectra, such as younger ages \citep{Whitler23}, `bursty' star formation rates that rapidly vary \citep{GSun23}, an evolving contribution of nebular continuum emission \citep{Katz24} and/or a top heavy IMF that has relatively more massive stars \citep{Cueto24}. Finally, it is also possible that there are additional sources of UV light that may become important at higher redshift, such as AGNs. Several of these scenarios can be addressed directly with more spectroscopic follow-up observations, but also by testing models for their consistency with redshift evolution of the shape of the UV luminosity function, or other observables such as the build-up of the galaxy stellar mass function \citep[e.g.][]{Donnan24}.

Spectroscopic redshift confirmations at $z>10$ remain challenging as typical spectra only detect the Lyman-break \citep{CurtisLake23}. The redshift obtained by such a broad feature, combined with the spectral resolution of the NIRSpec Prism mode and uncertainties in the modeling of damping wings from neutral gas in the intergalactic medium and dense gas in the interstellar medium \citep{Heintz24}, is not very precise. Notable are the detection of emission-lines in the rest-frame UV in deeper spectra, such as N{\sc iv},C{\sc iv}, He{\sc ii}, O{\sc iii}] and C{\sc iii}], as showcased by the spectrum of GN-z11 (whose detection in {\it HST} data was the first insight of a bright cosmic down beyond $z>10$; \citep{Oesch16,Bunker23,Maiolino24}). However, those UV emission-lines are often faint and their strength varies significantly among the population \citep{Harikane25}. The strong H$\alpha$ and [O{\sc iii}] lines have been detected by {\it JWST}/MIRI spectra of some of the most luminous galaxies \cite{Calabro24,Alvarez25}, but due to the lack of a multiplex it is challenging to obtain these measurements for statistical samples. Finally, ALMA has remarkably detected far infrared emission-lines in a galaxy at $z=14$ \citep{Schouws25,Carniani25}, which provides unique insights into the properties of their interstellar medium.

\subsubsection{(Bursty) star formation histories} \label{sec:bursty}
The star formation histories of galaxies are sensitive to physical processes such as gas inflows from the intergalactic medium, the formation and destruction of molecular clouds, dynamical instabilities and environmental effects such as galaxy mergers \citep[e.g.][]{ForsterSchreiber20}. Since these processes act on different time-scales, their relative importance impacts how rapidly star formation rates in galaxies vary \citep[e.g.][]{Tacchella20}. Due to the short age of the Universe, the light that we observe from early galaxies is mostly sensitive to processes that act on relatively short time-scales, compared to galaxies in the later Universe. Young and rapidly varying (`bursty') star formation histories have significant impact on the interpretation of galaxy spectra as light from the youngest stellar populations could outshine older, more massive ones \citep[e.g.][]{Gimenez23}, \citep[c.f.][]{Cochrane25}. 

Significant emphasis in the recent literature has been on characterizing the variability of star formation rates on $\sim1-100$ Myr time-scales in galaxies at $z\sim4-7$ \citep[e.g.][]{Asada24,Caputi24,Dressler24,Cole25,diCesare25}. At these redshifts, {\it JWST} can sensitively capture various features that are sensitive to such variations, such as the stellar UV continuum, nebular line and continuum emission and Balmer breaks. Before {\it JWST}, the presence of strong nebular emission-lines as [O{\sc iii}] and H$\alpha$ was inferred from their impact on broad-band photometry measured by {\it Spitzer}/IRAC \citep{Smit14}. Combined with ground-based data, observations indicated that the typical equivalent width of the galaxy population increases significantly with increasing redshift \citep{Stark16} and with decreasing mass \citep{Reddy18}. These trends have been confirmed for relatively bright galaxies by {\it JWST} spectroscopy \citep{Matthee23,Boyett24}.

More surprisingly, there are also indications that there is a population of galaxies that have experienced a downturn in their recent star formation rate. Such galaxies are increasingly more prevalent among the population of faint galaxies and can be identified through their relatively low H$\alpha$/UV ratios \citep{Endsley25}. Several high-redshift galaxies have been confirmed with spectra that display a relatively blue UV continuum (displaying the presence of relatively young stars), but without nebular lines \citep{Strait23,Looser24,Looser25}. These galaxies have been called `napping'/`mini-quenched'/in a lull phase/`smoldering' (the most applicable name yet to emerge) and while they are young, they did not form stars in the last few million years that would power nebular emission-lines \citep{McClymont25}. So far, systematically characterising the prevalence of such phases and their dependence on global properties of galaxies such as their mass has been challenging spectroscopically due to the difficulty in detecting continuum breaks. Nevertheless, based on photometric data, including data taken with medium-bands sensitive to (the absence of) line emission (such as Figure $\ref{fig:sed}$), the first sample studies are being undertaken and are expected to provide insights into the physics of star formation and feedback at early times \citep{CoveloPaz25,Looser25,Trussler25}.

\begin{figure}
    \centering
    \includegraphics[width=0.98\linewidth]{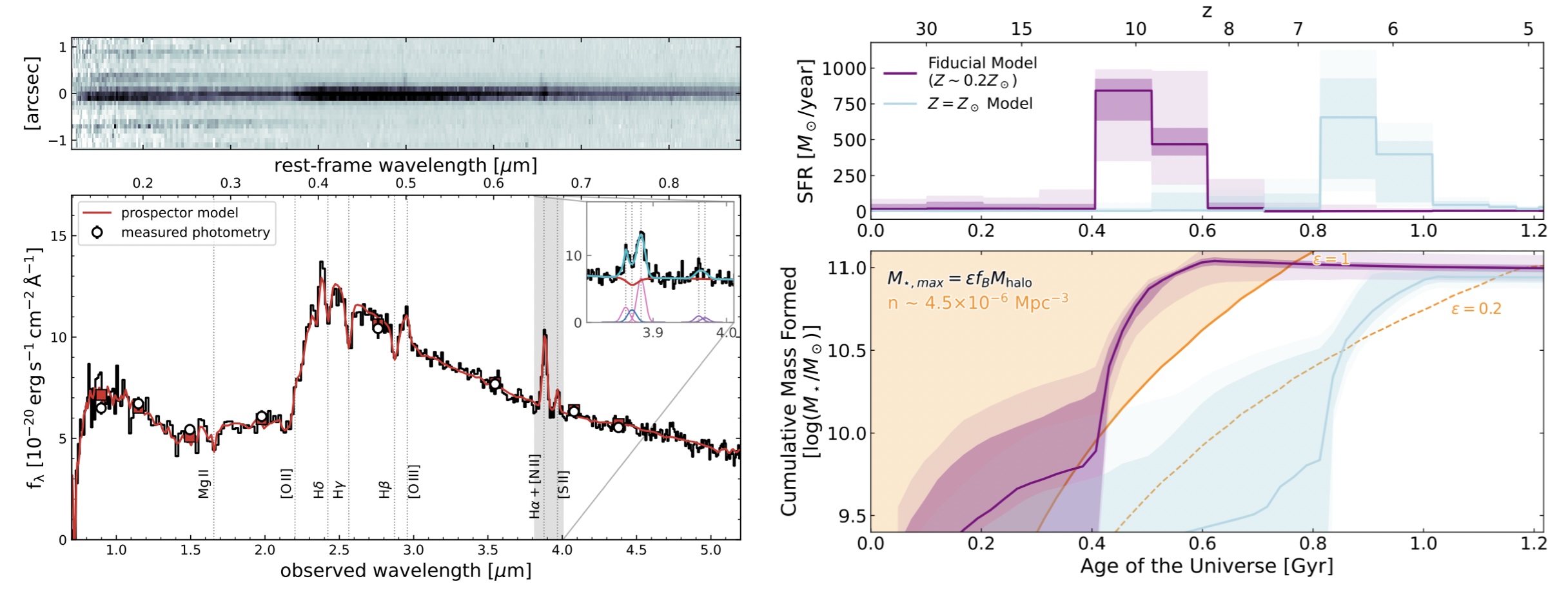}
    \caption{Left: a JWST/NIRSpec PRISM spectrum of a massive quenched galaxy at $z=4.9$ and a fit using stellar population models. The spectrum is characterized by a strong Balmer break and various absorption features from the Balmer series (H$\gamma$, H$\beta$), but also shows [N{\sc ii}] line-emission. Right: the modeled star formation history (top) and cumulative stellar mass history (bottom) of the galaxy, either with a solar (blue) or sub-solar metallicity (purple). The solid orange line shows the maximum stellar mass that could have formed for a typical halo given the number density of quenched galaxies at $z\sim5$ a formation efficiency $\epsilon=1$ and a universal baryon fraction. The dashed line shows $\epsilon=0.2$ which is the average value at the peak of the stellar mass - halo mass relation. The orange shaded region requires formation efficiencies above $\epsilon>1$, which are unphysical given standard assumptions. Adapted from de Graaff et al. 2025 \citep{deGraaff24}. }
    \label{fig:highzQG}
\end{figure}

\subsubsection{Rapid formation and quenching} \label{sec:highz_quench}

The study of the most massive galaxies in the early Universe constrains models of structure formation in the extreme regime. At a given age of the Universe, the assembly history of dark matter halos that can be calculated for a given cosmological model effectively yields an upper bound to the galaxy stellar mass that can assemble at a given time. 

A population of massive galaxies that {\it JWST} has been identifying and confirmed at $z\sim5$ are strongly obscured, highly star forming galaxies \citep[e.g.][]{Xiao24} that are typically `dark' (undetected) in the optical and near-infrared, but they are typically bright in the redder infrared bands and the sub-mm \citep{Williams19,Smail21}. Several of these galaxies exhibit complex and disturbed morphologies indicative of an active assembly through (multiple) galaxy mergers \citep{Barrufet23}. Likewise, these sources are reported to inhabit the most over-dense regions in the early Universe, such as proto-clusters \citep{Sun24overd}, in which a significant fraction of all the star formation in the early Universe took place.

Of particular interest are galaxies with Balmer breaks in their spectra, as those indicate that relatively old stellar populations already existed (Figure $\ref{fig:starsed}$). Since their star formation has quenched for a while, these galaxies are also used to study the causes of quenching. While passive galaxies had already been spectroscopically confirmed at $z\sim3$ before {\it JWST} \citep[e.g.][]{Schreiber18}, JWST spectroscopy allows the measurement of faint continuum and absorption line features with far higher precision and out to higher redshifts \citep[e.g.][]{Carnall23}. The sensitive photometric band coverage by NIRCam is improving our ability to perform statistical analyses and obtain robust estimates of their number densities \citep{Valentino23}, which are needed to compare to models of structure formation.

Early interpretations of the first {\it JWST} data indicated a high space density of candidate massive galaxies at very high-redshifts \citep{Labbe23Nat}, which could significantly challenge models of structure formation (these galaxies were the so-called Universe Breakers). Several early candidate massive galaxies turned out to be at lower redshifts, or have a significant AGN contribution to the rest-frame optical light (\cite{BWang24}, see Section $\ref{sec:LRDs}$). However, a high number of quenched galaxies with strong Balmer breaks are being confirmed spectroscopically at increasingly higher redshifts \citep{Glazebrook24,deGraaff24}, now out to $z\sim7$ \citep{Weibel25}. These spectra facilitate the reconstruction of the galaxies' star formation histories, as illustrated in Figure $\ref{fig:highzQG}$. In this figure, the spectrum on the left shows the clearly distinct double break detected at $z=4.9$, as well as various Balmer-series absorption lines (e.g. H$\gamma$, H$\delta$), together indicative of an evolved stellar population with ample A stars. The right panel of Figure $\ref{fig:highzQG}$ shows reconstructed star formation histories and the cumulative stellar mass assembly history of the galaxy using state-of-the-art SED fitting models that use flexible star formation histories. These assembly histories are combined with simple formation models that follow the evolution of dark matter halos with the same number density as observed for the passive galaxy in question, for different so-called `formation efficiencies', $\epsilon$. An efficiency above $>1$ would imply that there are more baryons in the galaxy than the average baryon fraction of the Universe, or that the dark matter halo mass would be significantly higher (at odds with the number densities). The mass assembly history of this galaxy is close to or even above the typical maximum efficiency ($\epsilon=1$), but it depends on the metallicity, assumptions on the star formation history and other variables \citep[e.g.][]{Carnall24}.

High-redshift quenched galaxies are also interesting testbeds for astrophysics such as the cause of quenching \citep[e.g.][]{Baker25}. A fraction of high-redshift quenched galaxies shows evidence for current or past AGN activity, suggestive of a connection to SMBH growth, feedback and quenching. For example, the galaxy in Figure $\ref{fig:highzQG}$ shows narrow [O{\sc iii}] and [N{\sc ii}] emission-lines that are indicative of shock ionization from an AGN, and other sources are reported with very broad Balmer lines originating from the broad-line region of an AGN \citep{Carnall23,Ito25}.

\subsection{Cosmic reionization} 
The CMB radiation originates from the last scattering of photons, which occurred when the Universe's expansion led to a low enough temperature for free electrons and protons to recombine into neutral hydrogen. The majority of the intergalactic baryons remained neutral for a few hundred million years, until the hydrogen atoms were reionized by ionizing photons emitted from early galaxies. Cosmic reionization is the story about how the first ionizing photons that are produced in young stars and accretion disks around (supermassive) black holes escape their natal clouds and ionize gas in the IGM. The conclusion of cosmic reionization significantly impacted the formation of low-mass galaxies by suppressing gas cooling onto small halos \citep{Sawala16}. In the following, we will discuss the latest insights on the timeline and sources that drove cosmic reionization primarily driven by new sensitive spectroscopy with {\it JWST}.

\subsubsection{The reionization timeline}  \label{sec:reion}
Directly measuring the progress of cosmic reionization by mapping the evolution of the 21 cm radiation from neutral gas is one of the key goals for upcoming radio telescopes such as the Square Kilometre Array \citep{Chapman22}. Currently, the two major constraints on the reionization timeline are obtained from the CMB \citep{Planck2020} and from the spectra of distant quasars \citep{Fan06}.

Free electrons from hydrogen that has been reionized scatter CMB photons through Thomson scattering. This has an impact on the large-scale polarization signal that yields a constraint on the total optical depth due to reionization and thus the average reionization redshift. The most recent constraint from the Planck satellite implies an average reionization redshift of $z=7.7$ \citep{Planck2020}. 

The end stages of reionization are currently best probed through the Gunn-Peterson trough and other Lyman-$\alpha$ forest statistics that are routinely measured in the spectra of quasars with redshifts $z>6$. Figure $\ref{fig:quasar}$ shows the rest-frame UV spectrum of a luminous quasar that covers the Ly$\alpha$ forest, the Gunn-Peterson trough, the broad Ly$\alpha$ line of the quasar and metal absorption series from foreground galaxies. Statistical measurements of the evolution of the transmission of the Ly$\alpha$ forest suggest that the end-stages of reionization were patchy \citep{Becker15Patchy} and ended at $z\approx5.3$ \citep{Bosman22}. Figure $\ref{fig:quasar}$ also marks the distribution of galaxies in the foreground, vicinity and background of the quasar. Various galaxy over-densities are seen, such as the over-density associated with the quasar, but also over-densities that are associated with Ly$\alpha$ transmission spikes (e.g. at $z=6.0$). The figure illustrates that large ionized regions traced by a relatively high Ly$\alpha$ transmission appear to be correlated with over-densities of galaxies \citep{Kashino25,Kakiichi25}.

\begin{figure*}
\begin{tabular}{cc}
    \centering
        \includegraphics[width=7.5cm]{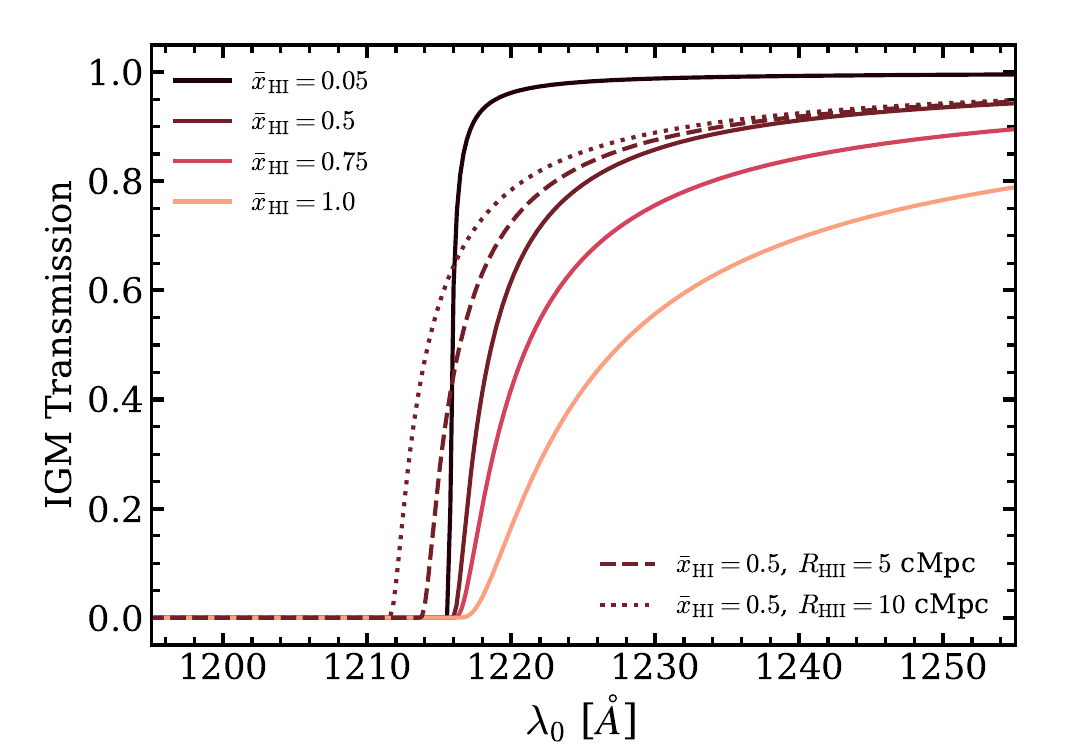} & 
   \hspace{-0.5cm} \includegraphics[width=6.5cm]{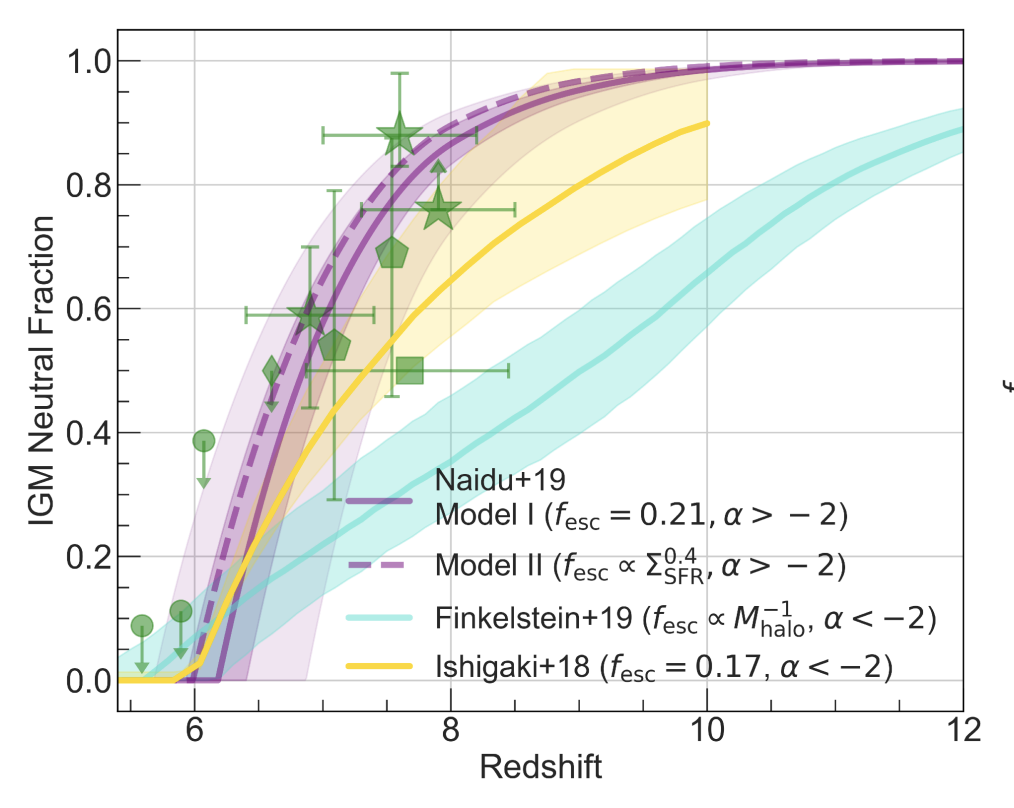} \\ 
    \end{tabular}
\caption{Left: illustration of the Ly$\alpha$ damping wing transmission for a range of neutral fractions and ionized bubble sizes. The damping wing extends to larger observed wavelengths in case the neutral fraction is higher, whereas the ionized bubble size primarily sets the sharpness of the damping feature \citep[e.g.][]{Witstok25a}. Right: the evolution of the neutral fraction from various measurements and the dependence of the reionization history on the main sources that dominated the ionizing emissivity (compilation and models from Naidu et al. 2020 \cite{Naidu20}, reproduced by permission of the AAS). If reionization were primarily driven by relatively rare, luminous sources, the history was relatively late and rapid, whereas an emissivity dominated by low mass halos would have yielded a reionization history that starts earlier and progresses more gradually.}
    \label{fig:eor}
\end{figure*}

As the Ly$\alpha$ absorption saturates above volumetric neutral fractions of $\sim10^{-4}$, it only probes the end stages of reionization. Currently, detailed measurements of the earlier part of reionization require other techniques based on galaxy or quasar spectra. One such technique uses the Ly$\alpha$ damping wing. 
The shape of the Ly$\alpha$ damping wing is sensitive to the neutral fraction of the intergalactic gas along that sight-line, as illustrated in the left panel of Figure $\ref{fig:eor}$. So far, several damping wing measurements have been obtained in luminous quasars beyond $z>7$ \citep{Banados18,FWang20,Durovcikova25}, and on (averages of) galaxy spectra \citep[e.g.][]{Umeda24,Mason25}. The main current challenges are the high required sensitivity, the modeling of the underlying stellar and/or quasar continua \citep[e.g.][]{Greig22}, damped Ly$\alpha$ absorption from gas clouds in galaxies \citep{Heintz24} and accounting for biases due to galaxies and quasars residing in large-scale over-densities \citep[e.g.][]{Kist25}.

The progress of reionization can also be mapped using observed statistics of the emerging Ly$\alpha$ emission-line from galaxies, such as their escape fraction and line-profile \citep[e.g.][]{Hayes23}. The observed Ly$\alpha$ line profiles are similarly impacted by the Ly$\alpha$ damping wing and are more easily observed than faint continuum features. Early estimates of the evolution of the neutral fraction have been made using Ly$\alpha$ demographics of distant galaxies \citep{Tang24,Kageura25}. Various studies report a connection between large-scale over-densities and the observability of Ly$\alpha$ emission, effectively extending the connection between galaxy over-densities and ionized regions out to $z\sim9$ \citep[e.g.][]{Saxena23,Whitler24,Witstok24}. Current efforts are pushing these measurements to the highest redshifts \citep{Witstok25} and aiming at separating the relative impacts of bright and faint galaxies in their connection to ionised bubbles \citep[e.g.][]{Torralba24}. Moreover, as the Ly$\alpha$ escape fraction and line-profile are also strongly sensitive to the interstellar gas conditions in galaxies, a detailed understanding of the correlations among those properties is required to make Ly$\alpha$ a reliable tracer of the progress of reionization \citep[e.g.][]{Lin24}.

\subsubsection{The sources of reionization} 
The two main types of sources of (re)ionizing photons are young stars in galaxies and AGNs. Galaxies are several orders of magnitude more numerous than UV-luminous AGNs (quasars), however, quasars are also much more luminous and may have a higher escape fraction of ionizing photons. Under standard estimates of their (escaping) ionizing luminosities, galaxies appear to dominate the ionizing photon emissivity, especially at early times \citep[e.g.][]{Dayal25}. 

As illustrated in the right panel of Figure $\ref{fig:eor}$, the timing of reionization depends on the luminosity of the sources that drove it \citep{Sharma18,Naidu20}, reflecting the later assembly of massive, luminous galaxies. The ionizing emissivity from the galaxy population depends on the shape of the galaxy luminosity function and the efficiencies with which ionizing photons are produced by stars and can escape the interstellar gas in galaxies \citep{Robertson15}. The galaxy ionizing output is parameterized as the product of $\xi_{ion}$, the ionizing photon production efficiency, and $f_{\rm esc}$, the escape fraction of ionizing photons. The key uncertainties are the number of faint galaxies beyond the detection limits and the luminosity dependence of $\xi_{ion} \times f_{\rm esc}$. A natural limiting luminosity would be the luminosity below which no galaxies exist, for example due to the atomic cooling limit \citep{BarkanaLoeb01}, M$_{\rm UV}\sim-13$. However, this cut-off has not yet been confirmed observationally \citep[see][for recent attempts]{Atek18,Kokorev25}.

While JWST enables progress in our understanding of the ionizing photon production efficiency \citep{Simmonds24,Llerena25,Pahl25}, the unknown escape fraction of ionizing photons $f_{\rm esc}$ remains the main bottleneck. Observations of ionizing photon escape in nearby galaxies \citep[e.g.][]{Jaskot24} point towards a complex picture where the escape fractions is correlated with the dust attenuation, the neutral gas distribution, the gas ionization parameter, the star formation rate surface density and the presence of galactic outflows \citep[e.g.][]{Flury25}. However, whether trends established in nearby galaxies apply to galaxies in the early Universe remains to be tested directly. The main redshift window to perform such tests is $z\sim2-3$, where the ionizing radiation is redshifted to $\sim400$ nm, so it can be observed for large samples with ground-based spectrographs as required to overcome the stochasticity of the foreground gas absorption \citep{Steidel18}. Studies have reported sample-averaged escape fractions of $\sim5$ \% \citep{Pahl21}, but these have so-far been limited to relatively bright galaxies, which do not capture the bulk of the galaxy population. 

A complementary approach to assess the role of galaxies in reionizing the Universe are studies that cross-correlate variations in the intergalactic gas properties at $z\sim5-6$ with galaxy over-densities \citep[e.g.][]{Kakiichi18,Meyer20,Kashino23}. This is particularly enabled by large samples of spectroscopically confirmed galaxies that JWST can deliver (see Figure $\ref{fig:quasar}$). A benefit of this approach is that the intergalactic gas is sensitive to all the ionizing photons, including those from undetected galaxies. In the epoch of reionization, the first JWST studies report excess ionization at distances $\sim5-50$ cMpc from galaxies, demonstrating that the sources that reionized the Universe are clustered with the detected galaxies \citep{Kashino25,Kakiichi25}. Due to saturation, Ly$\alpha$ transmission studies are limited to the end stages of reionization ($z\sim6$), such that future probes as the cross-correlation with the 21-cm signal are needed to extend studies into the reionization epoch.

\subsection{First stars and metal enrichment}
Hydrogen, helium and traces of lithium formed in the first twenty minutes of the Big Bang. The present-day Universe, however, has been enriched with a wide variety of elements that form the basis for molecules, chemistry and life to emerge. Metals\footnote{For astronomers, elements heavier than helium are called metals.} form during nuclear fusion that occurs inside stars and in the neutron-rich environments of supernova explosions. 

Identifying how the first stages of chemical enrichment occurred is among the key science cases for which JWST was envisioned \citep{Gardner06}. The metal enrichment history of the Universe can be investigated from multiple directions. Metal transitions from different elements and gas states can be detected in absorption in the interstellar medium \citep[e.g.][]{Sodini24}. On larger scales, metal absorption lines can be detected in absorption to bright background sources such as quasars, which probe metals around galaxies \citep[e.g.][]{Cooke11,Bordoloi24}. The presence of metals can be inferred from their impact on the dust attenuation in the interstellar medium \citep{Witstok23} and the re-emission in the infrared. Sensitive spectroscopy of (massive) stellar populations can also constrain the metallicity through their impact on photospheric absorption lines in stellar atmospheres and on the strength of P Cygni features originating from winds \citep{Steidel16,Chisholm19,Matthee22}. Finally, the gas-abundances in H{\sc ii} regions are studied through line-ratios of hydrogen and strong metal emission-lines such as [O{\sc iii}] and [N{\sc ii}], in particularly when combined with sensitive spectroscopy of auroral and recombination lines \citep{Kewley08,Andrews13,Curti24,Sanders24}. In this section, the main focus is on new insights on the metal abundances of H{\sc ii} regions, as this is where the majority of progress has been made with JWST so far.

\begin{figure}
\centering
    \includegraphics[width=13cm]{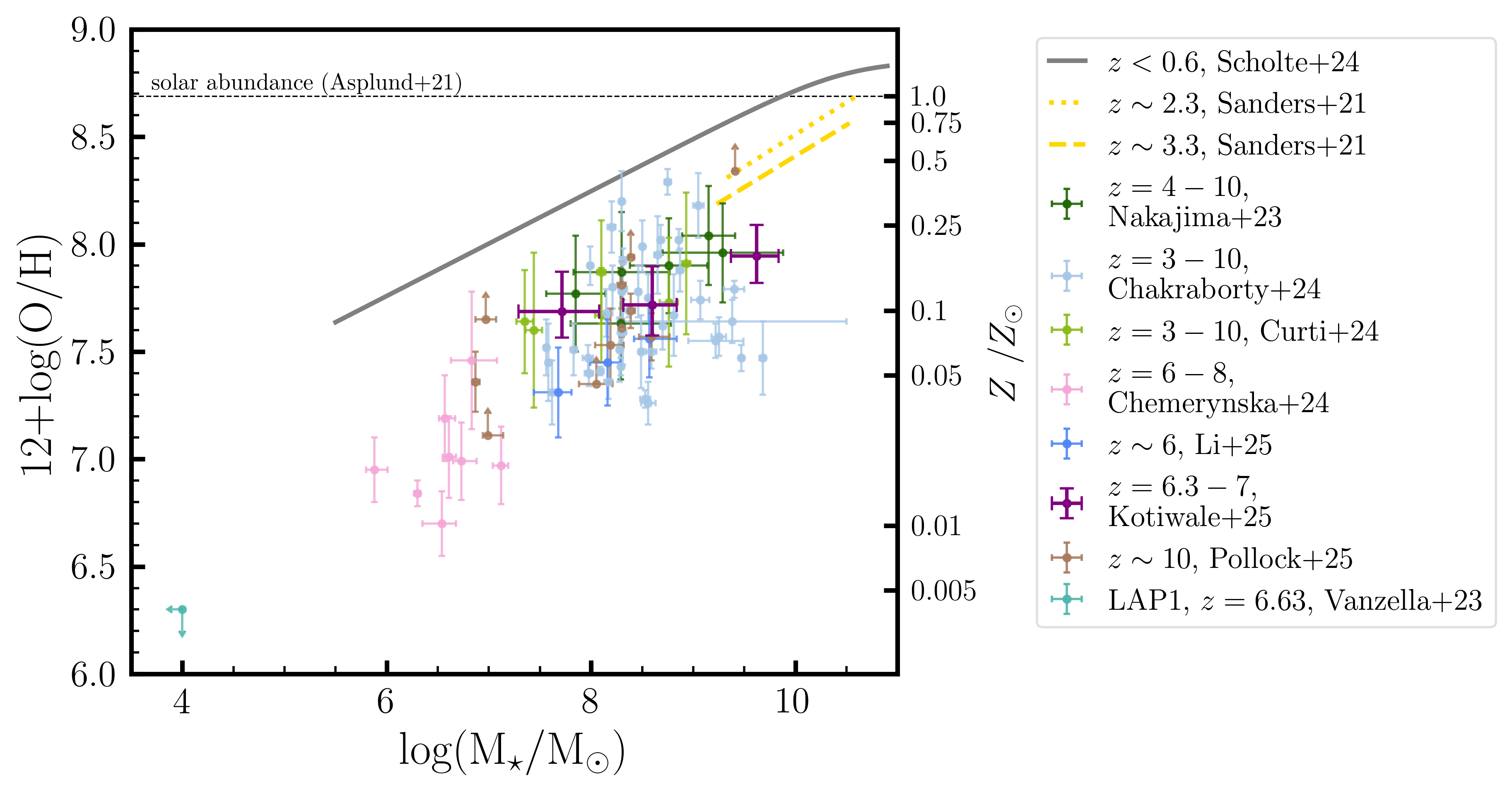} 
    \caption{Compilation of galaxy stellar mass and gas-phase metallicity measurements of galaxy populations at redshifts $z\sim0-9$ (adapted from Kotiwale et al. \citep{Kotiwale25}). The oxygen abundances of galaxies at high redshift are significantly lower than those in the local Universe \citep{Scholte24} and they extend relations probed at cosmic noon before JWST up to three orders of magnitude lower in mass. }
    \label{fig:mzr}
\end{figure}

\subsubsection{Metallicity and abundance measurements}
The study of gas-phase metal abundances in interstellar gas has a long historical record \citep{Tinsley79,Tremonti04,Andrews13} and was reviewed recently by Curti \citep{Curti25}. The variation of metal abundances within and among galaxies depends on the star formation history and the nucleosynthetic origins of the elements, but also on stellar winds and the accretion of newly accreted material \citep[e.g.][]{Kobayashi25}. The emission-lines that have traditionally been used to measure the gas-phase metallicity of galaxies are strong lines in the rest-frame optical. These strong lines trace common metals in the Universe (oxygen, neon, nitrogen) and are usually observed together with hydrogen emission lines from the Balmer series. Since oxygen is the most abundant metal and usually the strongest metal emission-line, the majority of gas-phase abundances are oxygen abundance measurements, especially at high redshift. This is an important distinction from measurements of metallicities in stars, which usually are iron abundances \citep{Steidel16,Strom22}. Significant challenges exist in the measurement of gas-phase metallicities because emission-line ratios are also sensitive to other properties of the gas, such as the ionization parameter, electron temperature, and electron density. 

Before the advent of JWST, the frontier of metallicity measurements for large statistical samples was at $z\sim2-3$ \citep{Sanders21}, because H$\alpha$ and [O{\sc iii}] redshift out of the $K$ band at $z\approx2.5, 3.5$, respectively. The main two advances brought by JWST are the extension of the redshift coverage of these emission-lines out to $z\sim10$ \citep{Pollock25} and, perhaps even more importantly, the strong sensitivity increase implies that much fainter emission-lines can be observed \citep{Strom23}. This is important to extend measurements of gas-phase metallicity to lower masses \citep{Chemerynska24}, but also to detect intrinsically faint auroral (such as [O{\sc iii}]$_{4364}$ and [O{\sc ii}]$_{7322,7332}$) and recombination lines that yield ‘direct’ metallicity measurements that can be used to calibrate relations between metallicity and strong-line diagnostics \citep{Nakajima23,Laseter24,Sanders24,Chakraborty24}. Nevertheless, these calibrations should be used with care. Unfortunately, the [O{\sc iii}]/H$\beta$ ratio -- which is relatively easily observable because the lines are strong and closely separated -- has a bimodal relation with metallicity, with a broad peak over the range 12+log(O/H) = 7.7-8.3, which coincides with the metallicity range of galaxies with masses $\sim10^{7-9}$ M$_{\odot}$ at $z\sim6$. Moreover, recent measurements of the electron density have yielded varying results suggesting that galaxies have stratified H{\sc ii} regions with a range in densities that impact the inferred properties of the ISM \citep{Reddy23,Ji24,Arellano25}.

Figure $\ref{fig:mzr}$ shows a compilation of measurements of the masses and gas-phase metallicites of galaxies at $z\approx0-10$ \citep{Scholte24,Sanders21,Nakajima23,Curti24,Chakraborty24,Chemerynska24,Li25,Pollock25}. The difference in the dynamic range in masses probed by JWST at high-redshift and by ground-based surveys that were previously probed at $z\sim2-3$ can clearly be noticed. Generally, the data suggest a mild evolution towards higher redshift. While $z\sim3-10$ measurements are clearly well below the $z=0$ relation, the measurements are primarily characterized by a large scatter. Apart from measurements from a handful of lensed galaxies \citep{Chemerynska24} that suggest a metallicity 12+log(O/H)$\approx7$ at masses $\sim10^7$ M$_{\odot}$ (and 12+log(O/H)$<6.5$ for a highly lensed arc; \citep{Vanzella23}), most of the JWST data at higher masses suggest a relatively flat trend between mass and metallicity relations \citep{Kotiwale25}. Such rapid enrichment could indicate a lower efficiency of stellar feedback with respect to lower redshift galaxies. Larger sample sizes with reliable metallicity measurements across a large dynamic range are required to more precisely measure the evolution of the mass - metallicity relation.

Rapid enrichment of metals is corroborated by the detection of metals out to the highest redshift galaxies known at $z\approx14$ \citep{Castellano24,Carniani24,Schouws25,Naidu25}. In the local Universe, galaxies are located on a relatively tight plane in the three dimensional SFR - stellar mass - gas-phase metallicity space, such that galaxies with a relatively high SFR have a relatively low metallicity. This relation is known as the `fundamental metallicity relation' \cite{LaraLopez10,Mannucci10} and it is usually interpreted in the context of galaxy regulator models \citep[e.g.][]{Lilly13}, where temporary increases in gas accretion lead to a lower metallicity and higher SFR. JWST measurements suggest that high-redshift galaxies appear systematically shifted towards lower metallicities than expected from the low-redshift 3D plane \citep{Heintz23,Curti24}. This could indicate that high-redshift galaxies are out of equilibrium, possibly due to an overflow of gas or due to bursty star formation that disturbs equilibrium timescales \citep{Torrey18,McClymont25}. However, galaxy selections usually are SFR limited, and not mass limited, especially at high-redshift when selections are based on strong emission-lines or the UV continuum. Whether this would bias the observed shape of the mass - metallicity relation is unclear and depends on the variability of galaxy star formation rates \citep{Kotiwale25}. Additionally, due to the young ages and bursty star formation histories, it is possible that only a few star forming regions in a galaxy are powering the observed emission-lines, such that their metallicities may not be representative of the full galaxy.

\begin{figure*}
\centering
    \includegraphics[width=14cm]{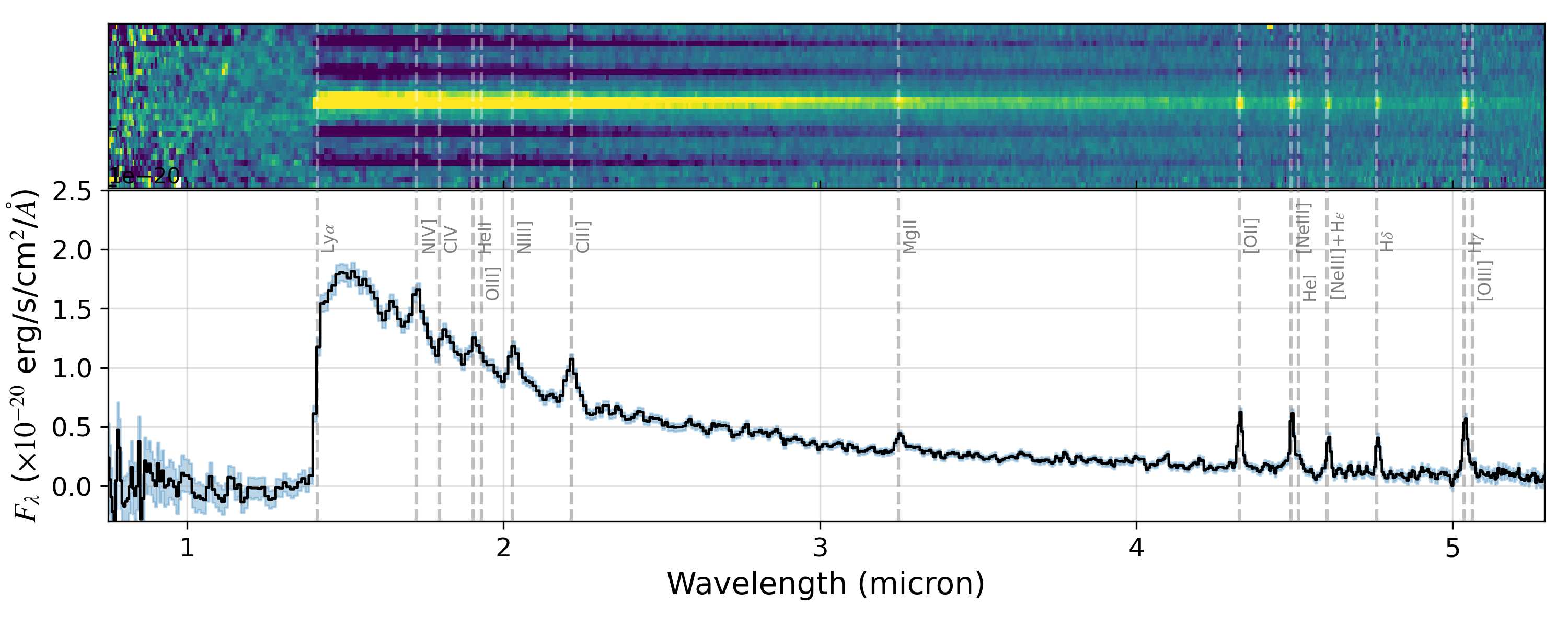}
    \caption{The JWST/NIRspec prism spectrum of the bright galaxy GN-z11 at $z=10.6$ \citep{Bunker23}. Besides the strong Lyman-break due to abundant neutral gas along the line of sight, the redshift is confirmed through the detection of a dozen of rest-UV to optical emission lines. The most unusual and discussed emission line among these suite of lines is the strong NIV]$_{1483,1486}$ doublet which relative strength to the oxygen lines indicates a dense nitrogen enhanced interstellar medium. }\label{fig:gnz11}
\end{figure*}

\subsubsection{The first stars and galaxies: expectations and early insights}
Historically, stars in the Milky Way has been classed as parts of Population I and II, where the first population of stars consists of young, metal rich stars mainly in the Milky Way's disk and the second population consists of older, metal poor stars that usually reside in the halo, in the bulge or in globular clusters. A hypothetical third population of stars (Population III, Pop III) are the first generations of stars that are composed exclusively of hydrogen and helium. One of the key goals of extra-galactic astrophysics is to identify these Pop III stars. As the majority of Pop III models suggest that these stars were likely very massive (due to less efficient fragmentation as a result of zero metallicity) and therefore short-lived, galaxies purely consisting of Pop III stars likely only existed at very high redshift and therefore were likely extremely faint \citep[e.g.][]{BarkanaLoeb01,Visbal15}. However, due to the possibly inhomogeneous nature of metal enrichment, it is possible that pockets of pristine gas survived out to later times\footnote{Indeed, the lowest metallicity known damped Lyman-$\alpha$ system has a metallicity of [Fe/H]$<-2.8$ \citep{Cooke11} at $z\sim3$, with C/O abundances consistent with the yields of a single Pop III supernova}, in theory enabling the formation of Pop III stars at lower redshifts \citep{Venditti23}. 

How would one identify a Pop III galaxy (or a clump of Pop III stars within a galaxy)? The cleanest detection would be the total absence of metals in the spectrum of a star/stellar population \footnote{In the Milky Way, stars with metallicities as low as [Z/H]$\approx-4.3$ have been reported \citep{Caffau11}.}, however, this would require extremely sensitive spectroscopy beyond our current capabilities (see \cite{Schauer20} for a discussion). More feasible strategies have either focused on identifying a very low gas-phase metallicity (mainly constrained through the oxygen abundance; e.g. \citep{Trussler23,Fujimoto25}) or on identifying expected signatures of Pop III stars such as very blue UV continua, unusual continuum shapes and strong He{\sc ii} line emission. The latter properties are in some way sensitive to the predicted hot temperatures of Pop III stars and its impact on nebular emission \citep{Zackrisson24}. The challenges associated with these techniques are that Pop III stars could be missed if the surrounding gas has been enriched from external sources, metal line emission could be quenched due to high densities (e.g. in case of the [O{\sc iii}]$_{5008}$ line) and that other ionising sources may mimic PopIII spectral signatures (i.e. He{\sc ii} can be powered by Pop II stars stripped in binary interactions \citep{Gotberg19}, or by AGN, shocks \citep{Lecroq24}).

In this context, recent JWST discoveries that highlight the current capabilities and interpretation-challenges, include: the record-low gas-phase abundance measured in a highly magnified galaxy at $z=6.6$ (oxygen abundance less than 0.5 \% solar; \cite{Vanzella23,Nakajima25}, see Figure $\ref{fig:mzr}$); an extremely blue, metal-poor galaxy at $z=8.3$ with very high effective temperature excessively heating the H{\sc ii} regions (a 1 \% solar oxygen abundance and an effective temperature of 80,000 K \cite{Cullen25}); and galaxies with highly unusual UV continuum shapes that could suggest the UV emission is dominated by the two-photon process (\citep{Cameron24,Katz24}, but see \cite{Tacchella24} for an AGN interpretation). Furthermore, Maiolino et al. \citep{Maiolino24PopIII} identified He{\sc ii} emission that could signpost Pop III stars in a clump associated to the luminous galaxy GN-z11, but without detections of hydrogen Balmer lines or UV emission at the same location, the interpretation of the origin of this line is challenging.

\begin{figure}[h]
\centering
    \includegraphics[width=10cm]{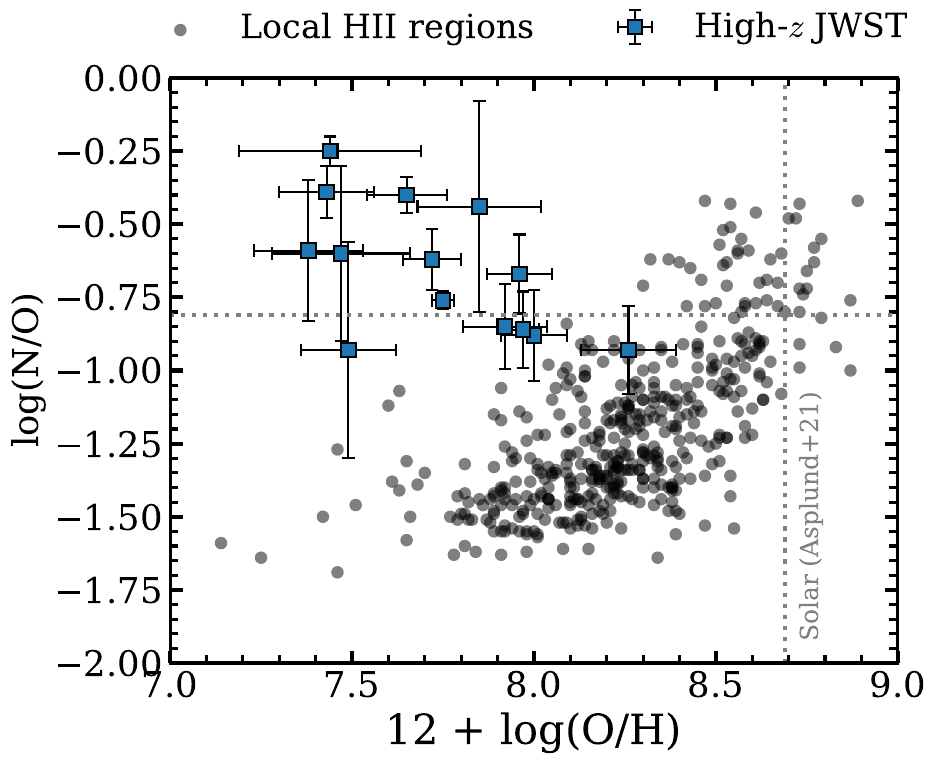} 
    \caption{A compilation of N/O abundances in the ISM of high-redshift galaxies and local H{\sc ii} regions. Blue squares show high-redshift galaxies as compiled by Ji et al. \citep{Ji25}. Grey circles show H{\sc ii} regions in the Milky Way from Berg et al. \citep{Berg20} and Izotov et al. \citep{Izotov06}. High-redshift galaxies are characterized by a nitrogen enhanced ISM with low oxygen abundance compared to H{\sc ii} regions.} 
    \label{fig:NO}
\end{figure}

Particularly noteworthy is the spectrum of the galaxy GN-z11 (Figure $\ref{fig:gnz11}$), which is one of the most distant galaxies that was found with HST due to its extra-ordinary brightness \citep{Oesch16}. The spectroscopic confirmation of the redshift GN-z11 of yielded both the first indication of the `bright cosmic dawn' discussed in Section $\ref{sec:earlySF}$ and the first example of a highly nitrogen-enriched galaxy in the early Universe \citep{Bunker23}. Such nitrogen enrichment has been found in various other high-redshift galaxies (\citep{Tang25}, see Figure $\ref{fig:NO}$) and caused significant debate regarding its origin, which could be linked to the formation of globular or nuclear star clusters, supermassive black holes or supermassive stars \citep{Senchyna24,Charbonnel23,Marques24,Antona25}. Like C/O abundances, N/O abundance measurements contains information on the specific episodes of metal enrichment \citep{Isobe23,Arellano24,Ji25,Naidu25} associated to phases of massive stellar evolution. Various (semi-) forbidden nitrogen and carbon lines that emit in the UV are sensitive to the electron density of H{\sc ii} regions and their unusual ratios suggestive of high electron densities have been among the key arguments why GN-z11 may host an AGN \citep{Maiolino24}. Now GN-z11 has demonstrated the capability of detecting various faint spectral features, it is clear that spectra with comparable sensitivity for much larger samples, extending to fainter magnitudes and higher redshifts can unveil new properties of the building blocks of the first galaxies.

\subsection{First supermassive black holes} 
Supermassive black holes (SMBHs) are present in virtually every massive galaxy in the local Universe \citep{KormendyHo13}. The feedback associated to their growth (AGN feedback) has become an indispensable ingredient in models of galaxy formation \citep{Weinberger18}, in particular, to quench the most massive galaxies (see \citep{Crain23} for a review). Massive galaxies have typically assembled most of their stellar mass early in the Universe (as JWST has witnessed; see Section $\ref{sec:highz_quench}$). Thus, it is likely that (some) galaxies also developed a significant SMBH mass at early times.

\subsubsection{Quasars: properties, environments and hosts}
Rapid SMBH growth has been detected in the high-redshift Universe already for decades, primarily through the high UV luminosity of quasars. Quasars have been detected out to redshifts $z\approx7.5$ \citep{Banados18,FWang21}, although they have very low number densities on the order of $10^{8-9}$ cMpc$^{-3}$ at $z>6$ \citep{Matsuoka18,Shen20}. High-redshift quasars are extremely luminous, reaching bolometric luminosities of $>10^{47}$ erg s$^{-1}$, and the SMBH masses implied from their luminosities and emission-line widths are in the range M$_{\rm BH}=10^{8-10}$ M$_{\odot}$ \citep{Yue24}. For a detailed review on the properties of quasars, including multi-wavelength observations of the gas around quasars, see the pre-JWST review by Fan, Ba{\~n}ados and Simcoe in 2023 \cite{FBS23}.

\begin{figure*}
\centering
\begin{tabular}{cc}
\includegraphics[width=7cm]{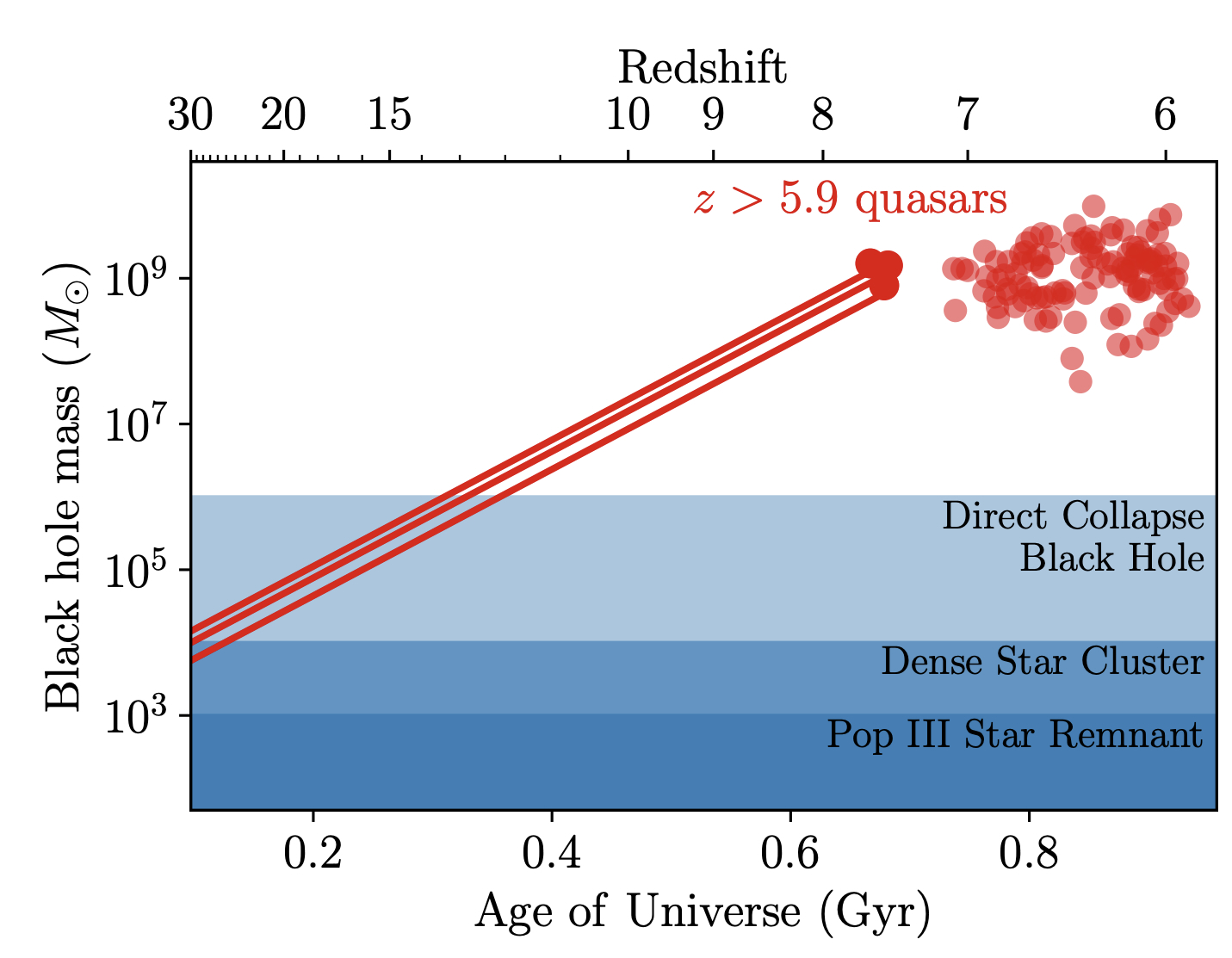} &
\includegraphics[width=7cm]{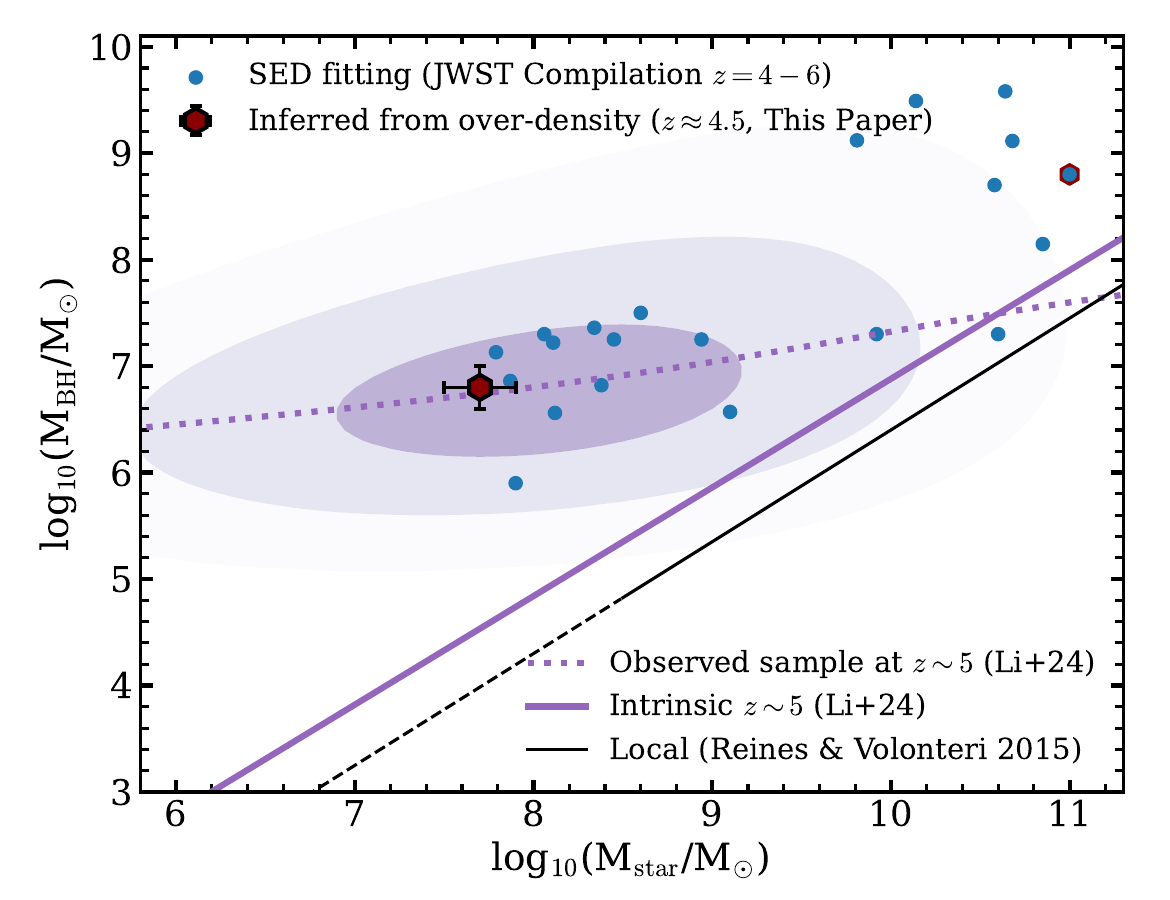} 
\end{tabular}
\caption{Left: The masses of high-redshift quasars (identified pre-JWST) versus redshift. The three curves show the exponential growth tracks for Eddington-limited accretion of three of the most massive quasars at $z>6$ in $\Lambda$CDM cosmology. The growth tracks suggest high `seed' masses $>1000$ M$_{\odot}$ (depending on the redshift of formation). Figure from \cite{FBS23}. Right: The relation between SMBH mass and galaxy stellar mass observed in high redshift quasars and AGNs, compared to the scaling relation found in the local Universe (black solid line; dashed in the extrapolated regime) and to a fit to the high-redshift data while modeling the bias that only the most actively accreting SMBHs at fixed stellar mass are likely observed (purple line, whereas purple contours show the expected distribution of observed AGNs). Figure from \cite{Matthee25}. } \label{fig:smbhs}
\end{figure*}

How could SMBHs in quasars have grown so massive so quickly? The left panel in Figure $\ref{fig:smbhs}$ illustrates this challenge by showing growth tracks of some of the highest redshift quasars with massive SMBHs assuming that they have been accreting at the Eddington limit. This limit corresponds to the accretion rate at which the radiation pressure generated by the accreting material equals the inward gravitational attraction, thus limiting further accretion. At the Eddington limit, black holes grow exponentially with the Salpeter e-folding timescale of $\approx5\times10^7$ yr. Nevertheless, depending on the redshift at which the SMBHs formed, Eddington-limited accretion suggests that these SMBHs had `seed masses' above $>1000$ M$_{\odot}$, i.e. well beyond the BH mass of the remnant of a massive star.

Possible pathways to assemble such massive SMBHs by $z\sim7$ include super-Eddington accretion that would enable faster growth \citep[e.g.][]{Husko24,Trinca24,Bennett24}, heavy $10^{4-6}$ M$_{\odot}$ seeds from a direct collapse of a large primordial gas cloud (without the intermediate step of fragmentation in stars and stellar evolution) \citep{Natarajan24,Bennett24}, or from the dynamical interactions and runaway collisions in dense stellar systems such as nuclear star clusters that lead to the formation of very massive stars that collapse into BHs with masses $\sim10^4$ M$_{\odot}$ \citep{PortegiesZwart2000,Zhu22}. However, definite evidence for any of these scenarios is lacking. It should be noted that SMBH masses are inferred indirectly through empirical calibrations based on quasars in the local Universe \citep{GreeneHo2005,Vestergaard09}. The first direct dynamical mass measurement of the SMBH mass in a quasar at $z\approx2$ hints at some systematic biases in the single-epoch virial mass estimators depending on the accretion rate \citep{Gravity24}, but these biases are within one order of magnitude. Clearly, more such measurements are warranted.

JWST observations of quasars have focused on probing the stellar masses of the systems hosting quasars. Despite the high resolution imaging data that JWST can provide, detecting host galaxies is extremely challenging due to outshining by quasars. For example, a typical quasar at $z=6$ has an AB magnitude of $\sim19$ at 2 micron, whereas the host galaxy would have a magnitude of $\sim25 (27)$ for a stellar mass of $10^{10 (9)}$ M$_{\odot}$. Careful subtraction of the quasar emission has revealed various indications of extended massive host galaxy detections (with masses $>10^{10}$ M$_{\odot}$; \citep[e.g.][]{Yue24}), but in some cases showing complex morphologies that indicate merger activity \citep[e.g.][]{Marshall24}. Spectroscopic studies of somewhat fainter quasars have directly identified stellar emission from the host galaxy in the rest-frame optical \citep{Ding23,Onoue24}, yielding more precise estimates of their stellar masses. Generally, despite the high inferred stellar masses, the BH to stellar mass ratios are on the order of $\sim1-10$ \%, demonstrating the efficient formation of SMBHs in the early Universe compared to the assembly of stellar mass. 

Independent insights into the systems that host quasars can be obtained from their large-scale clustering. Given their extremely low number densities ($\sim10^{-9}$ cMpc$^{-3}$; \citep{Schindler23}), one would naively expect that they are hosted by rare massive dark matter halos with masses $\sim10^{13}$ M$_{\odot}$. As the number densities of quasars are too low to perform quasar-quasar clustering measurements, the halo masses of quasars are inferred through their correlation with galaxies in the environment. JWST's NIRcam wide field slitless spectroscopic mode is highly efficient at mapping the large-scale environments on $\sim10$ cMpc scales in the plane of the sky \citep[e.g.][]{Kashino23}, see Section $\ref{sec:highz_jwstera}$. The average quasar resides in relatively large over-densities of galaxies, however, there appears to be a large variation in the galaxy over-densities around high-redshift quasars \citep{FWang23,Eilers24}. The average dark matter halo mass implied from the galaxy over-densities is $\approx3\times10^{12}$ M$_{\odot}$, which is similar to quasars seen at later epochs, i.e. $z\approx2$, \citep{Laurent17}. This is well below the halo mass that would correspond to the observed number densities of quasars. This implies that luminous quasars at $z\approx6$ do not reside in the most massive halos that exist at that cosmic time, but also that the duty cycle, which is the fraction of such halos that are detectable as UV-luminous quasars at a given epoch, is low. 

The low duty cycle indicates that quasar activity fluctuates rapidly and/or that most of the SMBH growth happens in an obscured phase. The latter is also supported by the relatively small proximity zones (i.e. ionized regions between the quasar and our line of sight mainly powered by the escaping UV radiation of the quasar) that have been measured in quasar spectra at $z>6$ \citep{Eilers20,Durovcikova25,Onorato25}. These small proximity zones have now also been confirmed along the plane of the sky in the spectra of background galaxies \citep{Eilers25} and suggest that the duration of the emission of ionizing radiation is relatively short and therefore imply that the majority of SMBH growth was obscured.

\subsubsection{Faint AGN \& little red dots} \label{sec:LRDs}
By being able to identify AGN activity spectroscopically, JWST has significantly extended samples of (faint) AGNs at $z>3$, probing up to three orders of magnitude lower luminosities compared to quasar samples \citep{Adamo24,Scholtz23}. Moreover, JWST data extended the redshift frontier to $z\approx10$ \citep{Tripodi24,Taylor25,Maiolino24}, well beyond the redshift of the most distant UV-selected quasars known.

\begin{figure*}[h!]
\centering
\includegraphics[width=14cm]{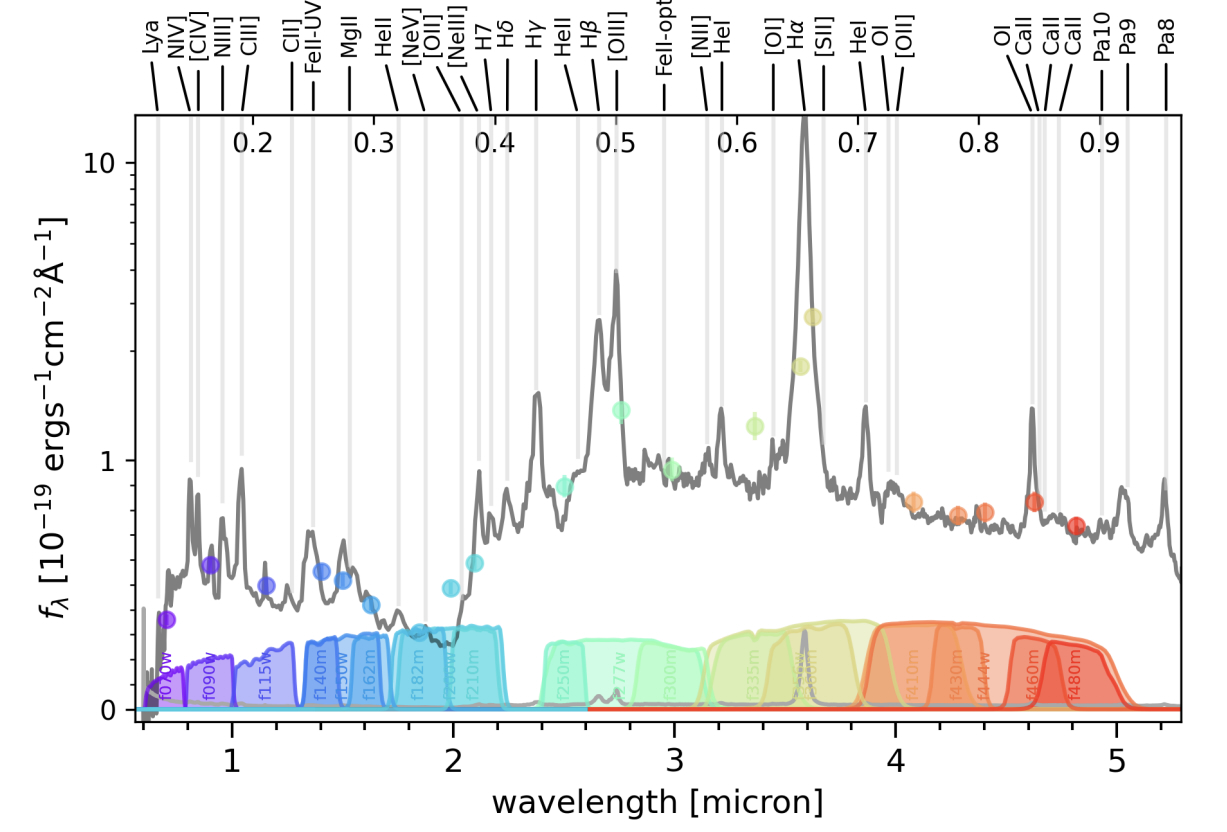} \\
\includegraphics[width=14cm]{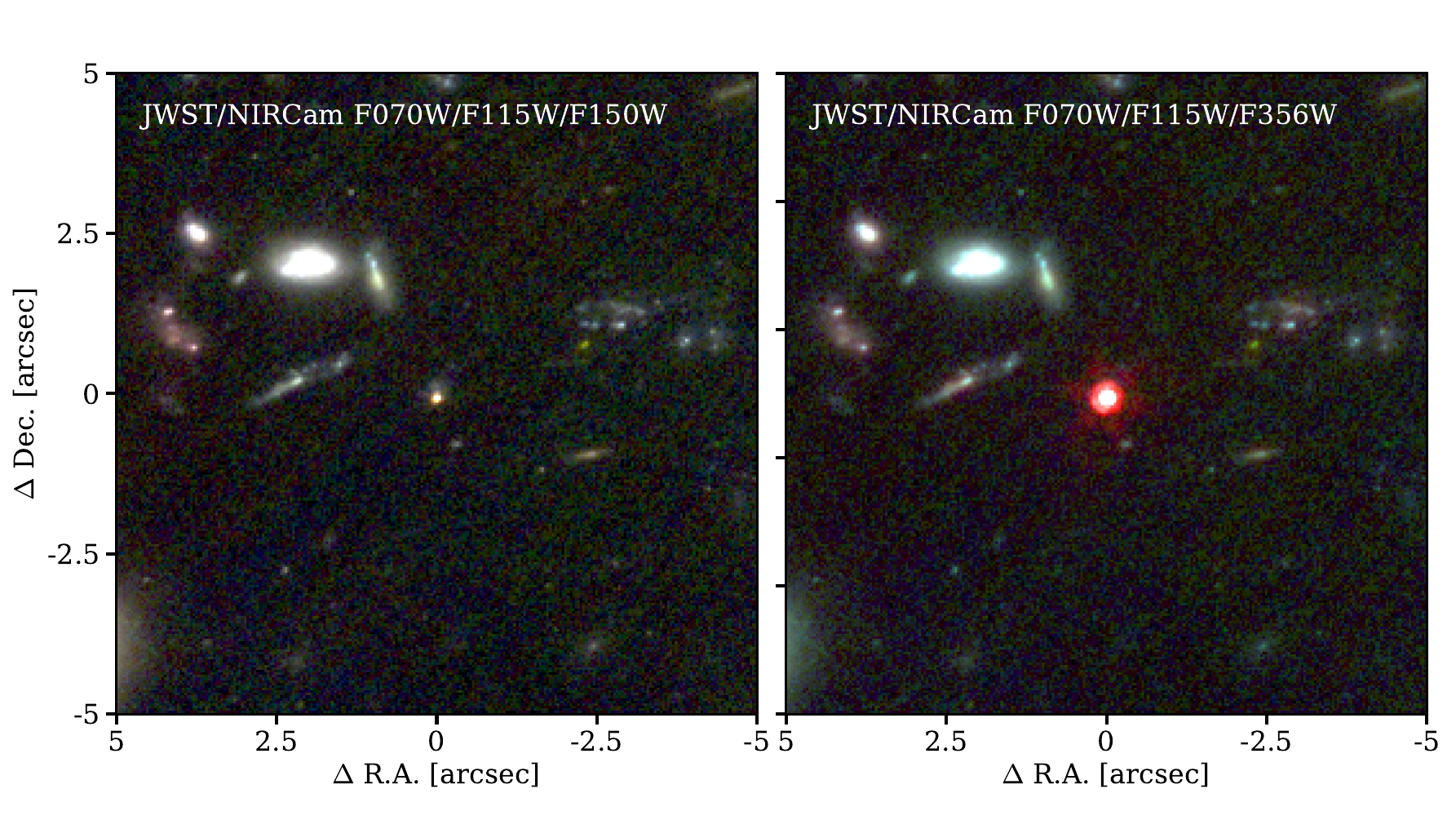} 
\caption{{\bf Top:} A deep JWST/NIRspec PRISM spectrum of the bright LRD A2744-45924 at $z=4.46$ (figure adopted from Labbe et al. \cite{Labbe24}). For reference, the NIRCam filter curves are added in the bottom. Vertical lines highlight the wavelengths of various emission features. The spectrum highlights the strong Balmer break (here redshifted to around 2 micron), strong and broad H$\alpha$ emission, high density gas tracers as Fe{\sc ii} and O{\sc i} as well as high excitation lines in the UV (such as N{\sc iv}] and C{\sc iv}). {\bf Bottom:} Zoom-in false color images of A2744-45924 based on JWST data. The left image uses filters in the blue end of JWST's coverage, which overlaps with filters previously available with HST, and that cover the rest-frame UV emission. The right image uses the same filters for the blue and green bands, but uses the redder F356W image in the red, which covers the rest-frame optical and the H$\alpha$ emission. In the left panel, the LRD does not particularly stand out compared to other galaxies, but in the right panel the LRD stands out for its extreme compact redness.} \label{fig:lrds} %
\end{figure*}

Among these AGN samples are the widely debated “Little Red Dots" \citep[LRDs;][]{Matthee24}. These constitute a loosely defined population of AGNs with a compact morphology and red UV to optical colors and broad Balmer lines (see the top panel in Figure $\ref{fig:lrds}$, \citep{Labbe23,Greene24}). LRDs have been found primarily at $z\approx4-6$ \citep{Kocevski23,Harikane23,Kokorev24,Matthee24,Lin24LRD,Greene24} where their number densities appear to peak, with a mild decline at higher redshifts. While LRDs have now been found at $z\approx0$ \citep{Izotov08,Lin25lowz,Ji25_Lord}, their number densities in the local Universe appear to be very low, extending the declining number densities with increasing cosmic time reported at redshifts $z<4$ \citep{Kocevski25,YMa25}.

Due to their red UV to optical continuum colours (see the top panel in Figure $\ref{fig:lrds}$), LRDs have been identified in photometric selections of various different types of galaxies, such as massive (quenched) galaxies at early times and highly obscured star-forming galaxies \citep{Barro23,PerezGonzalez24}. Indeed, initial LRD discoveries were interpreted as extremely massive galaxies at high-redshift \citep{Labbe23Nat}, potentially challenging the cosmological paradigm due to their high inferred stellar mass. These stellar mass estimates are significantly lowered when accounting for AGN emission \citep[e.g.][]{Maiolino23b,Chen24,BWang24}.

Despite broad Balmer lines and other AGN signatures (such as very high ionisation lines or high density tracers as Fe{\sc ii} or O{\sc i}, \citep{Labbe24,Tripodi25,Torralba25,DEugenio25Irony}), the LRDs have highly remarkable AGN properties. The vast majority are undetected in X-Rays \citep{Ananna24,Yue24b}, with limits well beyond the expected X-Ray luminosities  given their H$\alpha$ luminosity \citep{Sacchi25}. LRDs also lack the typically strong dust continuum emission originating from the hot dust torus around the broad line region \citep{Williams24,Akins24}. These observations imply that the bolometric luminosities of LRDs are significantly lower than would be expected based on standard AGN models \citep{Greene25}. 

\begin{figure*}[h!]
\centering
\includegraphics[width=11cm]{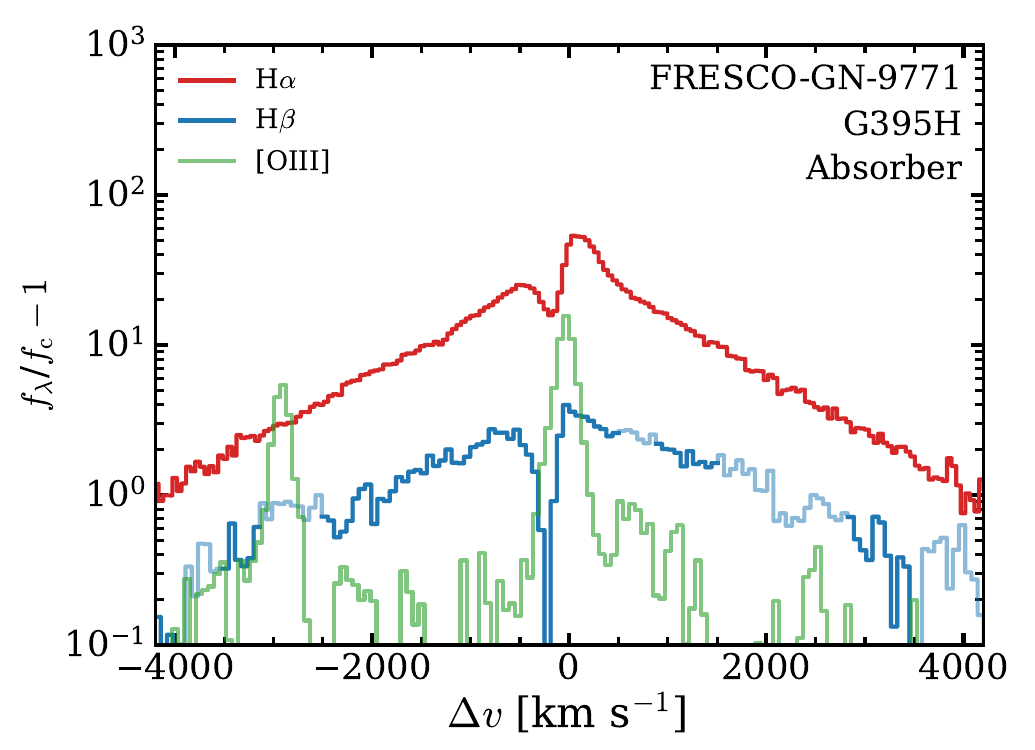} 
\caption{A sensitive, high resolution spectrum of a luminous LRD at $z=5.5$. The spectra are continuum-normalised to highlight differences in equivalent width. Shown are the H$\alpha$ (red), H$\beta$ (blue) and [O{\sc iii}] lines (green). The logarithmic scale highlights the exponential nature of the wings of the broad components of H$\alpha$ and H$\beta$. Broad components are absent in [O{\sc iii}], as some faint emission just right-ward of [O{\sc iii}]$_{5008}$ is due to HeI \citep{Torralba25}. Absorption is present in both Balmer lines, and stronger in H$\beta$ than in H$\alpha$.  
} \label{fig:lrd_Profiles} %
\end{figure*}

A significant fraction of the LRDs shows strong Balmer continuum breaks \citep{Setton24b}, unlike the power-law like spectra of quasars or reddened power-laws of obscured AGNs. Moreover, a high fraction ($\gtrsim30$ \%) has (relatively) narrow absorption lines in the Balmer series close to the systemic redshift, suggesting absorption by a high column of dense gas (see Figure $\ref{fig:lrd_Profiles}$,  \citep[e.g.][]{Matthee24,Juodzbalis24b,dEugenio25}). The absorption is mostly blue-shifted, but red-shifts have also been reported. The Balmer absorption is only seen in $<1 $\% of quasars \citep{Aoki06,Schulze18} and indicates a high covering fraction of dense excited hydrogen gas. The emerging spectrum of an AGN covered by such dense gas can produce the strong and smooth Balmer breaks that have been measured \citep{Inayoshi24,Ji25a,deGraaff25Cliff,Naidu25BH*}. Some attractive features of this configuration are that it may explain the X-Ray weakness due to the gas being Compton thick and it mitigates the need for a high dust attenuation, relaxing the lack of re-radiated dust continuum emission in the infrared and sub-millimeter \citep[e.g.][]{Akins25,Setton25,Xiao25}. Incorporating such models suggests that the galaxies hosting LRDs have relatively low stellar masses ($\sim10^8$ M$_{\odot}$; see Figure $\ref{fig:smbhs}$), which is supported by evidence from the relative weak clustering of faint AGNs with galaxies in their large-scale environments \citep{Matthee25,Lin25env}, dynamical masses \citep{Deugenio25b}, as well as their high number densities \citep{Pizzati24b}.

Based on calibrations derived from galaxies in the local Universe, the Balmer line width and luminosity \citep{GreeneHo2005}, the implied BH masses range from $\sim5\times10^{6-9}$ M$_{\odot}$. Combined with their stellar mass estimates, this implies very high BH to stellar mass ratios of $>10$ \%, possibly even approaching 1:1 (see Figure $\ref{fig:smbhs}$). These mass ratios are much higher than those of SMBHs found in galaxies in the local Universe \citep{KormendyHo13,Reines13} and therefore point toward highly efficient SMBH growth compared to the stellar mass assembly of a galaxy. This could mean that the mechanisms that lead to high `seed masses' (such as direct collapse black holes) occur at high-redshift \citep{Pacucci23}, or that the early growth of SMBHs (frequently) exceeds the Eddington limit \citep{Schneider23,Lupi24,Husko24}. However, there are reasons to be careful in interpreting the BH masses that are based on significant extrapolations of calibrations. In the gas-enshrouded AGN model, various radiative scattering processes could impact the observed Balmer emission-line profiles (see Fig. $\ref{fig:lrd_Profiles}$ for an example). For example, electron scattering could broaden the wings significantly to exponential profiles, and resonant scattering may impact the cores of line profiles \citep{Chang25,Rusakov25,Naidu25BH*,Torralba25}. This may lead to an over-estimate of the SMBH mass by up to two orders of magnitude, compared to when the line-width was assumed to be dynamical. However, the degree of scattering is debated \citep[e.g.][]{Brazzini25} and it is uncertain whether the scaling relation is applicable at all. 

Thus, it seems that the LRDs are revealing a previously unknown phase of gas-enshrouded AGN formation that is most prevalent in the early Universe \citep{Inayoshi25,Pacucci25}. However, many things remain unclear. What is the origin of the dense gas? Is it an envelope being expelled from an early `quasi-star' phase that nurtures the rapid formation of supermassive black holes \citep{Begelman25} or is the high covering fraction an indicator of near-spherical accretion flows theoretically expected for the most rapid accretion regimes \citep{Liu25}? What are the masses of the black holes powering LRD emission? Are there nearby star clusters associated to these black holes? Which mechanisms facilitate the rapid formation of massive black holes and how does the growth of these black holes impact their host galaxy? With the rapid advances being made in this field, one thing is clear: we are about to learn something new.

\section{Outlook} \label{sec:outlook}
The goal of this review was to introduce readers to some important basics underlying the observational studies of galaxies in the early Universe, but also to sketch some historical context and insights in areas where a lot of the current activity of the community is focused. Here, we will finish by providing an outlook for developments in the immediate future, primarily guided by personal expectations.

The JWST has been operational for only three years, but it has revolutionized our understanding of the early Universe. Its performance has been exceeding expectations \citep{Rigby23}, but the Universe itself has also kept surprising astronomers. JWST data have highlighted new discovery spaces that can be applied to other data-sets. For example, the prominent nitrogen lines in the rest-frame UV spectrum of GN-z11 at $z=10.6$ have motivated investigations into the abundances and gas densities in galaxies and globular clusters in the local Universe.  JWST has unveiled a previously unknown population of faint AGN with highly unusual spectra and physical conditions. In retrospect, such AGNs appeared already present at $z\sim0.1$ in the SDSS and were studied in 2008 \citep{Izotov08}, but were not given that much attention because they are very rare in the local Universe and earlier spectra did not identify unusual features as the Balmer line absorption. It may also well be that various seemingly separate developments may turn out to be quite related. For example, nitrogen-enriched gas has been associated with (nuclear) star cluster formation that may be the sites in which early SMBH growth happens \citep[e.g.][]{Isobe25}, whereas the mechanisms that cause bursty star formation in early galaxies could also disturb equilibrium scaling relations in variations among metallicity and star formation rates \citep[e.g.][]{McClymont25NO}. 

In the next few years, detailed JWST follow-up studies will target the newly identified populations of galaxies, while prioritising wavelength ranges of special interest with very deep spectroscopy. Simultaneously, wide-area surveys that are starting or about to start, such as the optical Legacy Survey of Space and Time by the Vera C. Rubin Observatory and the sensitive wide-area survey in the near-infrared by the Euclid and Roman telescopes and their spectroscopic follow-up campaigns, are poised to identify new exciting targets such as low-redshift LRDs or very high-redshift quasars. The Extremely Large Telescope is also imminent, with first light currently planned by the end of this decade. Due to its large light gathering power and the possibility to design more complicated instruments, we can primarily expect significantly improved (rest-frame UV) spectra of galaxies in the distant Universe and images at higher spectral and spatial resolution. The Square Kilometre Array may detect the 21-cm signal from the dark ages, which will enable new cosmological and astrophysical constraints, particularly on the sources and timeline of reionization. Finally, the lessons learned from these observational facilities are being incorporated in new, more complete models of galaxy formation.

\section*{Acknowledgements}

I thank Claudia Di Cesare, Edoardo Iani, Gauri Kotiwale and Wendy Sun for proofreading, Daichi Kashino, Gauri Kotiwale, Sara Mascia, Benjam\'in Navarrete and Joris Witstok for their assistance in preparing some of the Figures, and Richard Ellis and Stephen Blundell for constructive comments.

\section*{Funding}

Funded by the European Union (ERC, AGENTS,  101076224). Views and opinions expressed are however those of the author(s) only and do not necessarily reflect those of the European Union or the European Research Council. Neither the European Union nor the granting authority can be held responsible for them.

\section*{Notes on contributor}

\begin{wrapfigure}{l}{0.25\textwidth}
  \centering
  \vspace{-2em}
  \begin{minipage}{0.23\textwidth}
    \includegraphics[width=\textwidth]{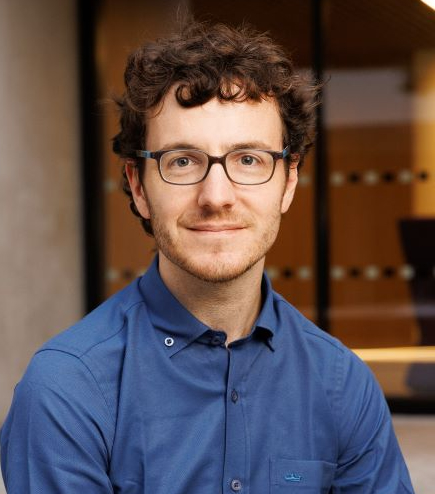}
  \end{minipage}
\end{wrapfigure}

Jorryt Matthee is an Assistant Professor at the Institute of Science and Technology Austria (ISTA), where he leads a research group on the astrophysics of galaxies. His work focuses on high-redshift galaxies, cosmic reionization and active galactic nuclei, primarily using observations from large telescopes. Matthee studied in the Netherlands (BSc in Utrecht and MSc in Leiden) and completed his PhD at Leiden University in 2018. He was a Zwicky Prize fellow at ETH Z\"urich before moving to ISTA in 2023.

\bibliographystyle{tfnlm}
\bibliography{main}

\end{document}